\documentclass[oneside,11pt,reqno]{amsart}

\pdfoutput=1 
\usepackage[margin=2cm]{geometry}    
\usepackage{amsaddr}

\usepackage{hyperref}
\usepackage{cite}
\usepackage[T1]{fontenc}
\usepackage[normalem]{ulem}
\usepackage{standalone}
\usepackage{graphicx}			
\usepackage{amsmath}
\usepackage{amssymb}
\usepackage{amsthm}
\usepackage[capitalize]{cleveref}
\usepackage{color}
\usepackage{mathtools}
\usepackage{tikz-cd}
\usepackage{xfrac}
\usepackage{bm}
\usepackage{bbm}
\usetikzlibrary{arrows}
\usetikzlibrary{decorations.markings}
\usetikzlibrary{shapes.geometric}
\usetikzlibrary{decorations.pathmorphing}
\tikzset{snake it/.style={decorate, decoration=snake}}
\tikzset{arrow/.style={postaction={decorate,decoration={markings,
                mark=at position 0.5 with {\arrow{>}}}}}}

\tikzset{end arrow/.style={postaction={decorate,decoration={markings,
                mark=at position 1 with {\arrow{>}}}}}}

\newcommand{\ket}[1]{\left| #1 \right>}

\newcommand{\eps}{\varepsilon}

\newcommand{\tr}{\mathrm{tr}}

\newcommand{\cA}{\mathcal{A}}
\newcommand{\cC}{\mathcal{C}}
\newcommand{\cD}{\mathcal{D}}

\newcommand{\cL}{\mathcal{L}}
\newcommand{\cO}{\mathcal{O}}

\newcommand{\cZ}{\mathcal{Z}}

\newcommand{\Z}{\mathbb{Z}}
\newcommand{\fT}{\mathfrak{T}}

\renewcommand{\Vec}{\mathsf{Vec}}
\newcommand{\Rep}{\mathsf{Rep}}

\newcommand{\Bsym}{\mathfrak{B}^{\text{sym}}}
\newcommand{\Bphys}{\mathfrak{B}^{\text{phys}}}

\newcommand{\phys}{\text{phys}}
\newcommand{\Neu}{\text{Neu}}
\newcommand{\Dir}{\text{Dir}}

\hypersetup{
    colorlinks=true,
    linkcolor=blue,
    citecolor=blue,
    filecolor=blue,
    urlcolor=blue,
}

\begin{document}
\setcounter{tocdepth}{1}

\title[{Reflections on time-reversal in SymTFT}]{Reflections on time-reversal in the Symmetry Topological Field Theory}
\author{\vspace{-0.2cm}Lea E.\ Bottini}
\address{\vspace{-0.2cm}\small Institut des Hautes Études Scientifiques, Bures-sur-Yvette, France}
\email{bottini@ihes.fr}
\author{\vspace{-0.2cm}Nick G. Jones}
 \address{\vspace{-0.2cm}\small St John’s College and Mathematical Institute, University of Oxford, UK}	
 \email{nick.jones@maths.ox.ac.uk}
 \thanks{The published version of this article is SciPost Phys. Core 9, 050 (2026); \href{https://scipost.org/SciPostPhysCore.9.3.050}{https://scipost.org/SciPostPhysCore.9.3.050}}
\begin{abstract}
Symmetry under time-reversal appears in the microscopic description of many physical systems. In a quantum mechanical setting it acts as an anti-unitary operator, so does not fall under general analyses based on unitary symmetries. In classifying zero temperature phases of matter in (1+1)d lattice models, the role of anti-unitary symmetries is, however, well-understood. In recent years, the Symmetry Topological Field Theory (SymTFT) approach to this classification has given a general framework to understand symmetries as topological defects, but does not naturally include anti-unitary symmetries. 
Following recent proposals in the literature, we adopt a symmetry-enriched SymTFT for a theory with both internal and time-reversal symmetry. In particular, we take a standard SymTFT associated with an internal unitary symmetry that is then enriched by a background time-reversal symmetry. A detailed analysis of the topological boundary conditions of this enriched SymTFT allows us to characterize the corresponding (1+1)d gapped phases that preserve the enriching symmetry (i.e.\ those that do not spontaneously break this symmetry in the ground state).
Line operators in the SymTFT approach are related to non-local string-order parameters (with charged end-point operators) for SPT phases. These are subtle in the anti-unitary case and we explore them both on the lattice and in the continuum. We include an analysis of unitary string order parameters that reveal the Klein bottle SPT invariant. On the lattice, we show that the correct end-point charge coincides with the time-reversal-charge only when the end-point operator is hermitian.
\end{abstract}
\maketitle
\tableofcontents
\section{Introduction}
Classifying zero temperature phases of matter and their transitions is a central problem in theoretical condensed matter physics. Without symmetries, different universality classes of quantum many-body systems are labelled by their topological order. Considerable understanding of this classification has been gained from both a physical and mathematical perspective \cite{Wen17,Sachdev23,Kong20,Johnson_Freyd22}. Symmetries constrain the space of allowed systems and lead to a richer phase diagram, with symmetry-enriched topological (SET) and symmetry-protected topological (SPT) orders \cite{Pollmann10,Pollmann12,Chen11,Schuch11,Fidkowski11,Chen13,Kapustin14}. 

Recent developments in understanding generalized\footnote{Also referred to as categorical or non-invertible symmetries in this context.} symmetries have led to a modern renewal of the \emph{Landau paradigm} that describes phases of matter and their transitions \cite{McGreevy_2023,Bhardwaj:2023fca,Lootens25,Chen25}. Broadly speaking, the conventional Landau paradigm classifies phases of matter by the subgroup of the global symmetry that remains unbroken. 
In the generalized framework, more exotic (symmetry-protected) topological phases can be characterized by the spontaneous breaking of suitable generalized symmetries.
These symmetries are identified with topological defects \cite{Gaiotto:2014kfa}, and are described by (higher) fusion categories. For  a review of these generalized symmetries see Refs.~ \cite{Bhardwaj:2023kri,Brennan:2023mmt,Luo:2023ive,Gomes:2023ahz,Shao:2023gho,Costa:2024wks,Schafer-Nameki:2023jdn,Kaidi:2026urc}.

A physically appealing framework to understand such generalized symmetries is provided by the symmetry topological field theory (SymTFT) \cite{Ji:2019jhk, Gaiotto:2020iye, Apruzzi:2021nmk, Freed:2022qnc}. Associated to a $d$-dimensional system with symmetry, the SymTFT is a $(d+1)$-dimensional topological field theory placed on a manifold with two $d$-dimensional boundaries. One is the \emph{symmetry boundary}, which is topological and encodes all of the symmetry information. The other is the \emph{physical boundary}, not necessarily topological, which describes the dynamics. The original physical system is recovered by compactifying the transverse direction to the boundaries. This has the advantage that the symmetry information is decoupled from the particular physical realization, and can be analyzed in isolation.

The SymTFT has proved to be an effective framework when the symmetry is internal (both for standard unitary group symmetries -- possibly anomalous -- and categorical symmetries). In particular, the SymTFT can be used to classify gapped phases with such a symmetry, providing a simple picture for the possible spontaneous symmetry breaking (SSB) patterns and order parameters \cite{Bhardwaj:2023ayw,Bhardwaj:2023fca,Bhardwaj:2023idu}. See also \cite{Thorngren:2019iar,kong2020algebraic,Chatterjee:2022tyg,Moradi:2022lqp,Chatterjee:2022jll,Bhardwaj:2023bbf,Bottini:2025hri,Bhardwaj:2024qrf,Aksoy:2025rmg} for recent applications of the SymTFT to the problem of classifying phases of matter. 
However, it is not at all clear how to apply the full framework when we include space-time symmetries, such as time-reversal---the focus of this work. Note that such space-time symmetries underlie many physical phenomena and it is therefore desirable that they should be treatable in the framework. Time-reversal in particular is frequently relevant in experimental settings, for example in the phenomenology of topological insulators \cite{Kane05,Hasan10}.

One obstruction to extending the usual SymTFT approach to space-time symmetries is that it is not clear how to gauge them\footnote{Although there have been developments in this direction \cite{Harlow:2023hjb,Susskind:2026moe}.}. 
Recent work by Pace, Aksoy and Lam introduced the notion of space-time symmetry-enriched SymTFT \cite{Pace:2024acq,Pace:2025hpb}. The idea is to analyze the internal symmetries as usual, and to then allow a non-trivial interplay between the space-time and internal symmetries. This is achieved by considering the standard SymTFT for the internal symmetry (where this is gauged) enriched by the space-time symmetry (treated as a background). This will necessarily miss some phases of matter---the enriching space-time symmetry remains unbroken---but does allow for a richer theory than in the case of internal symmetries only. We note that a SymTFT approach to classifying phases of matter with space-time symmetries has been recently adopted also in \cite{Antinucci:2025fjp,Apruzzi:2025hvs,Ravindran25,Orii26}. 

For simplicity, in this work we focus on the case of (1+1)d systems (and thus (2+1)d SymTFT), and on time-reversal symmetry. A key goal of our work is to explore the time-reversal-enriched SymTFT, and to expose some of its distinctive features. As a preliminary step, we analyze SymTFTs enriched by a unitary internal symmetry in order to build intuition for which gapped phases should be captured in the time-reversal-enriched case. In particular, we demonstrate for a variety of symmetry groups that this framework captures \emph{all phases} that do not spontaneously break the enriching symmetry (we consider only finite symmetries in this work). This restriction is structural rather than technical: since we do not have a prescription to gauge time-reversal, and so this remains as a background symmetry, the order parameter signaling its spontaneous breaking---a non-trivially charged anyon ending on both boundaries---is  unavailable. This restriction does not, however, prevent us from diagnosing phases in which time-reversal breaks together with an internal unitary symmetry, since the relevant order parameter is then provided by a (charged) anyon of the internal SymTFT. Provided time-reversal is not broken on its own, the enriched SymTFT reproduces the full classification in the examples we consider, and for the reasons given we expect this to be true in general.

There exist, of course, separate approaches to incorporating time-reversal, and other space-time symmetries, into the classification of phases of matter \cite{Pollmann10,Chen13,Kapustin14}. Indeed, it is known in (1+1)d that $G$-symmetric SPT orders are classified by $H^2_\epsilon(G,U(1))$, where $\epsilon$ indicates a twist that is discussed in detail below. 
Our aim is therefore not to find a new classification, or to uncover previously unknown (1+1)d phases. Rather, we build an explicit bridge between the lattice description in terms of matrix-product states (MPS) and the continuum SymTFT framework, clarifying several conceptual features of the time-reversal enriched SymTFT along the way. Instead of using space-time enrichment primarily to identify when a boundary fails to be symmetric, see e.g.\ \cite{Pace:2025hpb}, we characterize the symmetric gapped boundaries in more detail; for example, a single symmetric boundary of the internal SymTFT can split into several distinct boundaries once the time-reversal enrichment is accounted for, and it is this refinement that lets us identify order parameters built from unitary strings carrying non-trivial charge under the anti-unitary symmetry. We also clarify how to understand the apparently ambiguous end-point time-reversal charge of such a string order parameter. Although the anti-linear time-reversal charge is not gauge-invariant, we show that the appropriate transformation is an operator transposition (as well as conjugation by an on-site unitary), and that this reduces to the naive time-reversal charge precisely when the end-point operator is hermitian.

Establishing this bridge and deepening our understanding of how time-reversal can be incorporated into the SymTFT is highly desirable. This is because the SymTFT is a natural framework for treating generalized non-invertible symmetries, and the interplay of such symmetries with space-time symmetries is ripe for exploration, also in higher dimensions \cite{Antinucci25,Bhardwaj:2024qiv,Bhardwaj:2025jtf,Bhardwaj:2025piv,Wen:2025thg}. Note that the crystalline equivalence principle can reduce the classification problem to that of internal symmetries \cite{Thorngren18}, but we wish to work in a more general setting and see this result emerge.

The outline of the paper is as follows. In \Cref{sec:MPS}, we review the MPS approach to classifying symmetric phases of matter on the lattice, and how SPT phases can be revealed by string order parameters. Particularly interesting is the interplay between internal symmetries and time-reversal in this context. In \Cref{sec:symTFT} we review certain aspects of the usual SymTFT that will be useful going forwards, including the appearance of string-order parameters. This can be skipped by the expert reader. In \Cref{sec:timereversal} we discuss the relevant graded category needed in the case of time-reversal and explain symmetry-enrichment of SymTFT with a particular focus on the gapped boundaries. In \Cref{sec:examples} we then apply this to the classification of gapped phases for various symmetry groups. We first study (unitary) $\mathbb{Z}_4$ viewed as $\Z_2$ extended by $\Z_2$ to gain intuition. We then consider symmetry groups $\Z_2^T$, $\Z_n\rtimes\Z_2^T$ and $\Z_4^T$, where a superscript $T$ indicates the anti-unitary nature of the symmetry generator. We compare the approach to string orders in the SymTFT and through MPS throughout the text, and discuss exotic anti-unitary strings that are accessible only in the MPS setting in \Cref{app:anti-unitarystring}. Finally, in \Cref{sec:outlook}, we discuss future directions.

\section{(1+1)-dimensional lattice models and time-reversal symmetry} \label{sec:MPS}
In this work we focus on gapped quantum systems \cite{Zeng19}. Informally, this means that we have a finite set of degenerate ground states with all excited states separated by a finite energy gap. (In the TFT description of the space of ground states, this gap goes to infinity.)

We consider systems symmetric under some (non-anomalous and finite) symmetry group $G$, that will, in general, include anti-unitary (`time-reversing') elements. In particular, we have a map $\epsilon : G \rightarrow \mathbb{Z}_2$ that is 1 on unitary elements and $-1$ on anti-unitary elements. When we talk about invariants of a phase of matter, we mean quantities that are stable under all $G$-symmetric perturbations that do not close the gap. In lattice models, we have an on-site representation of this group \begin{align}g \rightarrow \left(\prod_{j \in \mathrm{sites}} u_j (g)\right) \mathcal{K}^{(1-\epsilon(g))/2}\ ,\end{align}
where $\mathcal{K}$ is complex conjugation in some fixed on-site basis\footnote{This fixed basis can be changed by incorporating a basis change into the operators $u_j$.}. 

This symmetry acts on the space of ground states. In the simplest cases, we have a unique vacuum and the symmetries preserve this vacuum state. We still have the possibility of non-trivial phases on this unique vacuum, since these symmetries may fractionalize, giving SPTs. In the case of multiple vacua, fixing a reference physical vacuum state\footnote{Such a state breaks some symmetries and obeys cluster decomposition. Symmetric ground states are formed by superpositions of such states with GHZ entanglement \cite{Zeng19}.}, we have an unbroken subgroup $H\subseteq G$ (note that $H$ maintains the $\epsilon$-grading). The action of $H$ on the reference vacuum gives an induced action of $G$ on the ground state space, where elements of the set $G/H$  permute the vacua. The unbroken symmetries may themselves fractionalize to form an $H$-SPT phase. In (1+1)d these statements can be made precise using the framework of MPS \cite{Chen11,Schuch11,Cirac21}, and are recovered, for unitary symmetries, in the SymTFT framework \cite{Bhardwaj:2023idu,Bhardwaj:2023fca,Huang:2023pyk,Qi25}.

In the remainder of this section, we focus on the fractionalized symmetries, and so fix $H=G$. The $G$-SPT phases are classified by group cohomology or certain cobordism groups \cite{Chen13,Kapustin14}. While these classifications differ in general dimensions, they coincide for the (1+1)d case that is our focus.

\subsection{MPS and gapped ground states of quantum lattice models}
In (1+1)d chains, translation-invariant MPS are states of the form
\begin{align}
\ket{\psi} = \sum_{j_1,\dots, j_L} \tr\left(\mathcal{A}_{j_1}\dots\mathcal{A}_{j_L} \right) \ket{j_1\dots j_L}\label{eq:MPS}  \ ,  
\end{align}
where, for fixed $j$, $\mathcal{A}_j$ is a $\chi\times\chi$ matrix \cite{Cirac21}. Fixing finite $\chi$ while taking the thermodynamic limit, these states obey the entanglement area law and can thus be used to approximate ground states of gapped quantum chains \cite{Hastings07,Dalzell19}. The classification problem of gapped ground states in (1+1)d can then be analyzed by classifying the phase diagram of MPS states \cite{Pollmann10,Chen11,Schuch11,Fidkowski11}. We consider the case where we have a unique MPS ground state; the discussion can be generalized to include SSB phases. Notice that the state in \cref{eq:MPS} is invariant under MPS-gauge transformations
$\mathcal{A}_j\rightarrow \mathcal{M}\mathcal{A}_j\mathcal{M}^{-1}$. Hence, while symmetries preserve the vacuum state, they can have a non-trivial action on the MPS tensor. This is the symmetry fractionalization. 
\subsection{Symmetry fractionalization and topological invariants}
Fix a group $G$ with unitary subgroup $G_0$ and anti-unitary elements comprising $G\setminus G_0$. Let  $\epsilon: G\rightarrow \Z_2 $ be $1$ on $G_0$ and $-1$ otherwise. 
In a translation-invariant setting, our symmetries transform the site-independent MPS tensor as follows:
\begin{align}
u(g)_{jk} \mathcal{A}_{k}&= e^{i \vartheta_g}V_g^\dagger \mathcal{A}_{j}V_g\qquad g\in G_0\nonumber\\
u(g)_{jk} \overline{\mathcal{A}}_{k} &= e^{i \vartheta_g}V_g^\dagger \mathcal{A}_{j}V_g^{} \qquad g \in G\setminus G_0 \ . \label{eq:symmfrac}
\end{align}
The topological classification corresponds to the (twisted) cohomology class of the projective representation to which these fractionalized symmetries give rise.
The projective representation is given by the 2-cocycle $b(g,h)$ where
\begin{align}
V_g V_h &= b(g,h) V_{gh} \qquad g\in G_0 \nonumber \\
V_g \overline{V}_h &= b(g,h) V_{gh} \qquad g \in G\setminus G_0 \ . \label{eq:corep}
\end{align}
Since we can freely replace $V_g \rightarrow e^{i \phi(g)} V_g$ in \cref{eq:symmfrac}, there is a gauge freedom in defining the representation. Invariant combinations of $b(g,h)$ must be preserved by the coboundary $e^{i \phi(g)}e^{\epsilon(g) i \phi(h)}e^{-i \phi(gh)}$; this 
leads to the invariant being given by the cohomology class of the 2-cocycle. I.e.\ we recover the classification $H^2_\epsilon(G,U(1))$, where $\epsilon$ tracks the `corepresentation' \eqref{eq:corep} \cite{Wigner59}.

We can distinguish different cohomology classes by combinations of the $b(g,h)$ that do not transform under the above gauge transformation. Examples are 
\begin{alignat}{3}
\varepsilon(g,h) &= \frac{b(g,h)}{b(h,g)} & \qquad\qquad& g,h\in G_0 \qquad \hphantom{10} [g,h]=0\nonumber\\
\theta(a) &= b(a,a) & \qquad&a \in G\setminus G_0 \qquad   \hphantom{i}a^2=1\nonumber\\
\kappa(a,g) &= \frac{b(a,g^{-1})b(g,g^{-1})}{b(g,a)} & &  \hspace{-.25cm}\begin{cases} a \in G\setminus G_0 \qquad g \in G_0\\   ag^{-1}=g a \end{cases} \ .
\end{alignat}
These are, respectively, the discrete torsion phase, the crosscap, and the Klein bottle invariants. In field theory, these can be interpreted in turn as partition functions over $T^2$ with twists, $\mathbb{R}P^2$ and the Klein bottle with twist \cite{Shiozaki16,inamura2021topological}.

\subsection{String order parameters}\label{sec:stringorder}
\begin{figure*}
    \centering
    \includegraphics[width=\textwidth]{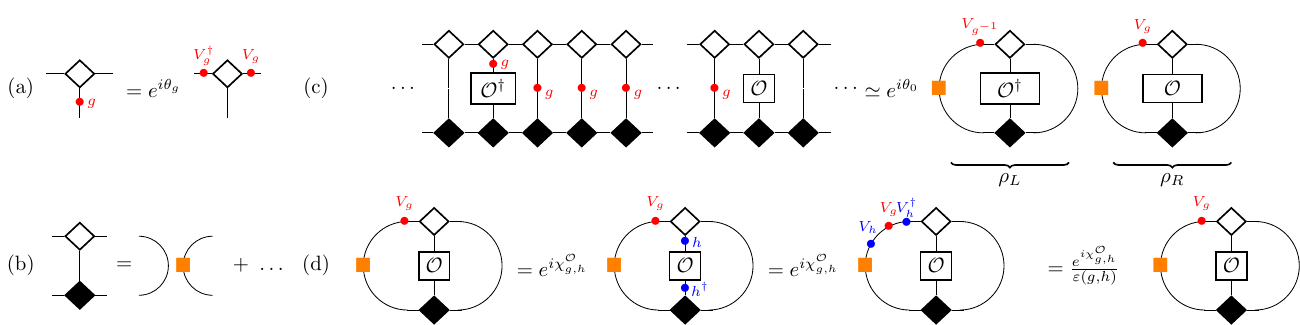}
    \caption{Graphical analysis of unitary string-order parameters in an MPS. The MPS tensor, in canonical form, is a hollow diamond, while its complex conjugate is a filled diamond. (a) Symmetry fractionalization of a unitary representation of $G$ on physical indices, $U(g) = \prod_j u_j(g)$. A transformation by $g$ is equivalent (up to a phase) to a projective representation $V_g$ acting adjointly on the bond indices. (b) The transfer matrix with unique dominant left and right eigenvectors. The orange square (an eigenvector containing the entanglement eigenvalues) commutes with $V_g$ for all $g$. (c) Define $\mu_k(g) = \prod_{j=-\infty}^{k-1}u_j(g) \mathcal{O}_k$; the two-point function $\langle \mu_1^\dagger(g) \mu_k^{}(g)\rangle$, for large $k$, factorizes as two local tensor terms $\rho_L$ and $\rho_R$. The phase $e^{i \theta_0}$ includes phases from the symmetry fractionalization, as well as from replacing $V_g^\dagger$ by $V_{g^{-1}}$, but does not play a role in any selection rules. (d) Consider $h$ such that $U(g)U(h)=U(h)U(g)$. The string order obeys a \emph{selection rule}. The local tensor term $\rho_R$ vanishes unless the charge $e^{i\chi_{g,h}^\cO}$ in $U(h) \mathcal{O} U(h)^\dagger = e^{i \chi_{g,h}^{\mathcal{O}}}\mathcal{O} $ is equal to $\eps(g,h)$, the discrete torsion phase of the projective representation. An identical condition holds for $\rho_L$.}
    \label{fig:unitarySO}
\end{figure*}

We will show how these invariants give selection rules for string order parameters. While the methodology we use is standard \cite{Pollmann12}, time-reversal symmetry introduces several subtleties, and we are not aware of such an analysis appearing elsewhere. String order parameters consist of a non-local \emph{unitary} symmetry string decorated by end-point operators charged under a second symmetry. The charge of the end-point is correlated with the symmmetry fractionalization of the first symmetry. Selection rules can be derived that show that certain topological invariants imply the vanishing of the string order parameter unless the end-point has a particular charge.  
In \Cref{app:anti-unitarystring} we explore the possibility of taking the non-local symmetry string to be anti-unitary---this is not a standard observable, but we show that, for MPS states, we can recover the crosscap invariant. More general order parameters have been analyzed for time-reversal symmetry, but they do not have the form studied here \cite{Pollmann12,Shiozaki16}; we discuss these in \cref{sec:outlook}. Note also Ref.~\cite{Liu23} for an approach to finding order parameters for time-reversal symmetric phases based on quantum convolutional neural networks; this work includes interesting no-go results for possible order parameters. 
\subsubsection{Unitary string order with end-point charged under a commuting unitary symmetry, discrete torsion}\label{sec:discretetorsion}
As a warm-up, let us first consider commuting unitary elements $g,h$. This is a case analyzed in \cite{Pollmann12}. We take a string operator for $g$, with end-point charged under $h$ (the charge is defined below). In particular, let us define $\mu_k(g) = \prod_{j=-\infty}^{k-1}u_j(g) \mathcal{O}_k$, where $\mathcal{O}$ is the end-point operator. The string-order parameter\footnote{More carefully, these are correlators of dressed symmetry strings that \emph{may} have long-range order. Their interpretation as an order parameter is discussed below.} is then the two-point function $\langle \mu_1^\dagger(g) \mu_k^{}(g)\rangle$.
Following the graphical reasoning of \cref{fig:unitarySO}, we end up with $\langle \mu_1^\dagger(g) \mu^{}_k(g)\rangle\simeq e^{i\theta_0}\rho_L\rho_R$, where we have a product of two tensor networks of the form
\begin{align}
\rho_R&= \sum \Lambda_\alpha^2 V_{g^{}}^{\alpha,\gamma}\overline{\mathcal{A}}_k^{\alpha,\beta} \mathcal{O}_{k,j}\mathcal{A}_j^{\gamma,\beta}\qquad\qquad
\rho_L = \sum \Lambda_\alpha^2 V_{g^{-1}}^{\alpha,\gamma}\overline{\mathcal{A}}_k^{\alpha,\beta} \mathcal{O}^\dagger_{k,j}\mathcal{A}_j^{\gamma,\beta} \ .\label{eq:twotensors}
\end{align}
We use the canonical form defined in Ref.~\cite{Cirac21}, and the approximate equality is asymptotic in string length, $k$, up to exponentially decaying terms. In \cref{eq:twotensors}, and similar equations below, we sum over all tensor indices.

Inserting $u(h)^{} u(h)^\dagger$, as in \Cref{fig:unitarySO}(d), on physical indices we find
\begin{align}
\rho_R&= \sum \Lambda_\alpha^2 (V_h^{} V_{g}V_h^\dagger)^{\alpha,\gamma}\overline{\mathcal{A}}_k^{\alpha,\beta} \tilde{\mathcal{O}}_{k,j}\mathcal{A}_j^{\gamma,\beta} \\&= \eps(g,h)^{-1}\sum \Lambda_\alpha^2V_g^{\alpha,\gamma}\overline{\mathcal{A}}_k^{\alpha,\beta} \tilde{\mathcal{O}}_{k,j}\mathcal{A}_j^{\gamma,\beta} \,, 
\end{align}
where $\tilde{\mathcal{O}} = u^{}(h) \mathcal{O} u(h)^\dagger $. Now, suppose that $\tilde{\mathcal{O}} = e^{i \chi_{g,h}^{\mathcal{O}}}{\mathcal{O}}$; then, $\rho_R = \eps(g,h)^{-1}e^{i \chi_{g,h}^{\mathcal{O}}}\rho_R$. Thus $\rho_R$ vanishes unless $e^{i \chi_{g,h}^{\mathcal{O}}}=\eps(g,h)$.

A similar calculation gives 
\begin{align}
\rho_L&= \sum \Lambda_\alpha^2 (V_h V_{g^{-1}}V_h^\dagger)^{\alpha,\gamma}\overline{\mathcal{A}}_k^{\alpha,\beta} \tilde{\mathcal{O}}^\dagger_{k,j}\mathcal{A}_j^{\gamma,\beta} \\&=  \eps(g^{-1},h)^{-1}\sum \Lambda_\alpha^2V_{g^{-1}}^{\alpha,\gamma}\overline{\mathcal{A}}_k^{\alpha,\beta} \tilde{\mathcal{O}}_{k,j}^\dagger\mathcal{A}_j^{\gamma,\beta} \,.
\end{align}
Now, we have that $\eps(g^{-1},h)=\eps(g,h)^{-1}$. Moreover, whenever $\tilde{\mathcal{O}}  =e^{i \chi_{g,h}^{\mathcal{O}}}{\mathcal{O}}$, we have $\tilde{\mathcal{O}}^\dagger  = e^{-i \chi_{g,h}^{\mathcal{O}}}{\mathcal{O}}^\dagger$. Hence, $\rho_L$ and $\rho_R$ both vanish unless $e^{i \chi_{g,h}^{\mathcal{O}}}=\eps(g,h)$. (Note, if this holds, neither need vanish but either might.) This selection rule, and the definite end-point charge, then allows us to use the non-vanishing of the string order to conclude that  $e^{i \chi_{g,h}}=\eps(g,h)$.

\subsubsection{Unitary string order with end-point `charged' under anti-unitary symmetry, Klein bottle}
In this section we study Klein bottle invariants and selection rules for $\mathbb{Z}_{2k}^T$. \label{sec:klein}
Let us consider a unitary string $g$ in the presence of an anti-unitary $\mathbb{Z}_{2k}^T$. Write the anti-unitary generator as $T=\left(\prod_j u(T)_j \right)\mathcal{K}$. This must have even order since the identity operator is unitary.

The symmetry-fractionalization of $T$ is
\begin{align}
u(T)_{jk} \overline{\mathcal{A}}_{k} = e^{i \vartheta_T}V_T^{\dagger}\mathcal{A}_{j} V_T^{} \ . \label{eq:Z2tfrac}
\end{align} 
Taking the conjugate we also have
\begin{align}
\overline{u(T)}_{jk} {\mathcal{A}}_{k} = e^{-i \vartheta_T}\overline{V}_T^{\dagger }\overline{\mathcal{A}}_{j}\overline{V}_T^{} \ .
\end{align}
Following again the graphical reasoning of \cref{fig:unitarySO}, our string order is asymptotic to a product of tensor networks of the form
\begin{align}
\rho_R&= \sum \Lambda_\alpha^2 V_{g^{}}^{\alpha,\gamma}\overline{\mathcal{A}}_k^{\alpha,\beta} \mathcal{O}_{k,j}\mathcal{A}_j^{\gamma,\beta}\qquad \qquad
\rho_L = \sum \Lambda_\alpha^2 V_{g^{-1}}^{\alpha,\gamma}\overline{\mathcal{A}}_k^{\alpha,\beta} \mathcal{O}^\dagger_{k,j}\mathcal{A}_j^{\gamma,\beta} \ .
\end{align}

We want to understand the interplay with the symmetry fractionalization of $T$, and by analogy with the discrete torsion case above we expect $\mathcal{O}$ to have a charge under this symmetry---we will return to the interpretation below. Let us simply repeat the analysis from above, which is complicated by the conjugate symmetry fractionalization.
We find
\begin{align}
\rho_R
&=\sum \Lambda_\alpha^2 (V_TV_{g}^t V_T^\dagger)^{\alpha,\gamma}\overline{\mathcal{A}}_{k}^{\alpha,\beta}\tilde{\mathcal{O}}_{k,j}^t A_{j}^{\gamma,\beta}\qquad\qquad
\rho_L
=\sum \Lambda_\alpha^2 (V_TV_{g^{-1}}^t V_T^\dagger)^{\alpha,\gamma}\overline{\mathcal{A}}_{k}^{\alpha,\beta}(\tilde{\mathcal{O}}_{k,j}^\dagger)^t A_{j}^{\gamma,\beta} \,,
\end{align}
where we transform operators by
$
  \tilde{\mathcal{O}}^t={u}(T) ^{}\mathcal{O}^t {u}(T)^\dagger$.
Assuming now that our symmetry forming the string, $g$, satisfies $Tg^{-1}=gT$, 
 we recognize the Klein bottle invariant $V_TV_{g}^t V_T^\dagger=\kappa(T,g) V_g$. Let us then suppose that we have a `charged' endpoint $\tilde{\mathcal{O}}^t= {u}(T) ^{}\mathcal{O}^t {u}(T)^\dagger=e^{i \phi_{g,T}^{\mathcal{O}}} \mathcal{O}$; we conclude that
 \begin{align}
\rho_R
=\kappa(T,g)e^{i \phi_{g,T}^{\mathcal{O}}}  \rho_R\qquad\qquad
\rho_L
=\kappa(T,g^{-1})e^{-i \phi_{g,T}^{\mathcal{O}}} \rho_L \ .
\end{align}
These give a single selection rule: the $g$-string order vanishes unless $\kappa(T,g)=e^{-i \phi_{g,T}^{\mathcal{O}}}  $.

In general, the physical meaning of the charge of the end-point is obscure. However, note that if we have a hermitian end-point operator then  
$\tilde{\mathcal{O}}^t= u(T)^{} \overline{\mathcal{O}}u(T)^\dagger = 
e^{i \phi_{g,T}^{\mathcal{O}}}  \mathcal{O}$. This is restrictive: a hermitian end-point necessitates that $e^{i \phi_{g,T}^{\mathcal{O}}} =\pm1$. Nevertheless, in this case the charge appears in the adjoint action of the anti-unitary symmetry $T$ on the end-point operator. If $g$ is a $\mathbb{Z}_2$ symmetry we can always choose such a hermitian end-point \cite{Verresen21} (see also \Cref{app:anti-unitarystring}). Moreover, $1=\kappa(T,1)=\kappa(T,g^2)=\kappa(T,g)^2$, so the charge of the hermitian end-point operator can be used to identify $\kappa(T,g)$. In more general settings it is useful to have the flexibility of non-hermitian end-points, but the charge is not then interpreted as an anti-unitary symmetry action.

In our analysis we did not make any assumption about the unitary symmetry $g$ that makes up the string. Suppose that $k$ is even---then $T^k$ is a unitary symmetry that squares to identity, satisfying $TT^{-k} = T^{k}T$. We hence have an invariant $\kappa(T,T^{k})$ that we can identify using the (genuine) charge of the hermitian end-point under $T$. Taking, for example, $k=2$, we have that $\Z_4^T$ has invariant\footnote{For $\Z_4^T$, note that $V_T\overline{V}_TV_T\overline{V}_T=\pm 1$ is clearly an invariant. By equating the different ways of simplifying this expression using \cref{eq:corep}, one can show this coincides with $\kappa(T,T^2)$.} $\kappa(T,T^2)=\pm1$, and we can detect the unique non-trivial SPT phase for $\mathbb{Z}_4^T$ using a string order parameter with a $T^2$-string.

For any value of $k$ we can detect non-trivial SPT phases whenever our total symmetry group contains a $\mathbb{Z}_N \rtimes \mathbb{Z}_{2k}^T$ subgroup. This corresponds to a group element $g$ of order $N$ that gives rise to a Klein bottle invariant.

\subsubsection{Discussion}
We can ask what practical use this analysis has---in the case we know the MPS and know the symmetry fractionalization, we can find the projective representation and identify the SPT phase. One benefit of traditional string order parameters, is that they can be defined without knowing the MPS description; we simply ask what a certain correlation function is. If it is non-zero, this means we cannot be in any of the phases that are inconsistent with this topological invariant. This does not tell us the SPT phase, rather it narrows it down (see Ref.~\cite{Pollmann12} for a unitary example where the discrete torsion charges are not enough to identify the phase). 

We contrast this with the case of `anti-unitary strings' that we analyse in \Cref{app:anti-unitarystring}. In this case, we cannot even define such an order parameter without writing down an MPS description. This is because we need the local structure to define the partial time-reversal (as in Ref.~\cite{Chen15}). Nevertheless, it is interesting that we can show that $\theta(T)$ can be recovered this way.

\section{SymTFT for internal symmetries}
\label{sec:symTFT}

\begin{figure}
\includegraphics[]{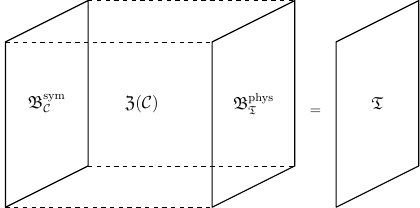}
\caption{The basic SymTFT sandwich: the $(1+1)$-dimensional theory $\fT$ on the right-hand-side is constructed as the interval compactification of a $(2+1)$-dimensional SymTFT $\mathfrak{Z}(\cC)$ on the left-hand-side, with two boundary conditions. The gapped (topological) boundary $\Bsym_{\cC}$ is on the left and the physical, possibly non-topological, boundary $\Bphys_{\fT}$ is on the right. If the physical boundary $\Bphys_\fT$ is also topological, the resulting theory $\fT$ is a $(1+1)$d TQFT.\label{fig:SymTFT}}
\end{figure}

In this section we review the SymTFT approach to the classification of gapped phases in $(1+1)$d following \cite{Bhardwaj:2023idu,Bhardwaj:2023fca}. The symmetries of a $(1+1)$d theory $\fT$ are described by a fusion category $\cC$, whose objects correspond to topological lines. $\cC$ is also called the \textit{symmetry category}. The SymTFT associated to $\fT$ is a  $(2+1)$d TQFT  and can be constructed as a gauged version of the symmetry $\cC$ in $(2+1)$ dimensions. The theory $\fT$ can then be recovered as the interval compactification of the SymTFT; this is  known as the sandwich construction, see \cref{fig:SymTFT}. 

The simplest example of SymTFT arises for group-like symmetries, where it is a (twisted) Dijkgraaf-Witten theory for the background gauge field of the symmetry and its dual. In general, we have no access to a simple BF-type action for the SymTFT, but we can still utilize the knowledge of its topological defects to extract certain key pieces of physical information. In particular, the topological defects of the SymTFT realize what is known mathematically as the Drinfeld center of $\cC$, denoted $\cZ(\cC)$.

The main idea is that gapped phases with a categorical symmetry $\cC$ are in one-to-one correspondence with topological/gapped boundary conditions of the SymTFT. These are in turn classified  by Lagrangian algebras $\cL$ in the Drinfeld center $\cZ(\cC)$. Algebraically, a Lagrangian algebra reads $\cL = \bigoplus_a n_a a$ and  specifies the simple anyons that are condensed at the boundary (clearly we have $n_1 = 1$). The coefficients $n_a \geq 0 $ are the  dimension of the vector space of local operators on which $a$ terminates on the boundary. For $\cL$ to be Lagrangian, these have to satisfy $\sum_a d_a n_a = \cD$, with $\cD = \sqrt{\sum_{b\in\cZ(\cC)}d_b^2}$ being the total quantum dimension of $\cZ(\cC)$\footnote{The coefficients $n_a$ must also satisfy $n_a n_b \leq \sum_c N_{ab}^c n_c$. So far, we specified $\cL$ only at the level of simple objects. Importantly, the definition of algebra includes also a multiplication morphism $\cL \otimes \cL \rightarrow \cL$. For a more complete definition, see \cite{kong2014anyon,Cong:2017ofj}. These details will not be important in our discussion.}. The anyons not in $\cL$ (which are confined to the boundary, as $\cL$ being Lagrangian implies that every anyon $a \notin \cL$ braids non-trivially with at least one anyon in $\cL$) are uncondensed and  become the topological lines generating a fusion category symmetry on the boundary.

\begin{figure}
\includegraphics[]{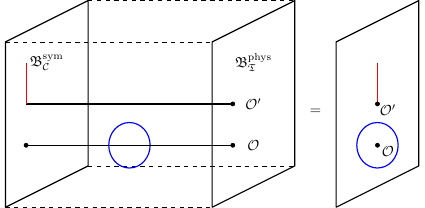}
\caption{Order parameters are characterized by anyons ending on the physical boundary $\Bphys$. Anyons also ending on $\Bsym$  give rise to local order parameters ($\cO$), while anyons becoming a symmetry line (red) on $\Bsym$ give rise to string-order parameters  ($\cO'$). The charge of an order parameter (obtained by linking a symmetry line (blue) with the operator) descends from the braiding properties of the corresponding bulk SymTFT anyons. Local order parameters $\cO$ are in 1-to-1 correspondence with ‘vacua’ $v$ of the resulting TQFT. While the $\cO$ diagonalize the symmetry action, with a change of basis we can map them to operators satisfying $v_a v_b = \delta_{ab} v_{a}$. These $v_a$ are the usual vacua, on which the symmetry acts by permutation \cite{Bhardwaj:2023idu}.\label{fig:sym_charges}}
\end{figure}

We fix the symmetry boundary $\Bsym_\cC$ to be specified by the Lagrangian algebra $\cL_\cC$, such that the symmetry realized on the boundary is precisely $\cC$. To obtain a gapped phase after `taking the sandwich' (i.e.\ doing an interval compactification, see \cref{fig:SymTFT}), for the physical boundary $\Bphys$ we \textit{also} fix a gapped boundary condition $\cL_\phys$. By cycling through all possible topological $\cL_\phys$, we span all gapped $\cC$ symmetric phases. In particular, this allows us to find (see \cref{fig:sym_charges}):
\begin{itemize}
    \item Order parameters (local and string)---these are non-vanishing operators in the IR that are low-energy limits of lattice operators with long range order. In this picture these correspond to anyons ending on the physical boundary, i.e.\ the anyons appearing in $\cL_\phys$. We explore the link to the lattice string order parameters in \Cref{sec:stringorder_SymTFT}.
    \item Vacua for the gapped phase that correspond to anyons completely ending on both boundaries. These give rise to topological local operators after taking the sandwich.
\end{itemize}

\subsection{The SymTFT where $\cC$ is a finite group}

We consider the case where the symmetry is given by a finite group $G$ (formally, this is described by the fusion category $\cC=\Vec_G$ of $G$-graded vector spaces). Then, as discussed in \Cref{sec:MPS}, the possible gapped phases are a mix of SPT and SSB phases. In particular, they are classified by a pair $(H,b)$, where $H$ represents an unbroken subgroup and $b \in H^2(H,U(1))$ is the SPT class. The broken vacua form a $G/H$ coset. 

This can be easily reproduced in the SymTFT setup. The first task is to construct the Drinfeld center $\cZ(G)$, whose anyons $a_{[g],r}$ in this case are labelled by a conjugacy class $[g]$ of $G$ and an irreducible representation (irrep) $r$ of the centralizer of a representative element $g \in [g]$. We can think of these as the topological lines of the $G$-gauge theory in $(2+1)$d, with $a_{[1],r}$ giving the charges in $\Rep(G)$ and $a_{[g],1}$ giving the $G$ fluxes. The remaining lines are of mixed charge/flux type.

Boundary conditions of the SymTFT correspond to Lagrangian algebras in $\cZ(G)$. The symmetry boundary  in this case is given by the Dirichlet boundary condition, corresponding to condensing all the charges 
\begin{align}
    \Bsym_G = \cL_\Dir = \bigoplus_{r \in \Rep(G)} \text{dim}(r)\,a_{[1],r} \,.
\end{align}
The flux lines $a_{[g],1}$ are instead given Neumann boundary conditions and become the $G$ symmetry generators on the boundary. 
Any other boundary condition is related to $\cL_\Dir$ by gauging a subgroup $H$ with discrete torsion $b \in H^2(H,U(1))$, which we will denote by $\cL_{\Neu(H),b}$. In particular, the fully Neumann boundary condition corresponds to condensing all the fluxes. In this case the charge lines $a_{[1],r}$ are free and generate the dual $\Rep(G)$ symmetry on the boundary. 

Choosing 
\begin{align}
    \Bphys = \cL_{\Neu(H),b}
\end{align}
gives a $G$-symmetric gapped phase after taking the sandwich, which obviously matches the classification given above. We mention some particular cases:
\begin{itemize}
    \item $\Bphys = \cL_\Dir$: SSB phase with no unbroken symmetry and $|G|$ vacua;
    \item $\Bphys = \cL_{\Neu(G),b}$: SPT phase for $G$ with a single vacuum;
    \item $\Bphys = \cL_{\Neu(H),b}$: SSB down to a subgroup $H$ with a corresponding $H$ SPT.
\end{itemize}

\subsection{SPT Phases from the SymTFT}\label{sec:standardSPT}
We consider now in greater detail how SPT phases are detected in the SymTFT via the paradigmatic example of $G=\Z_2\times\Z_2 = \{ 1,s,c,v \}$. Here we use $s,c$ for the first and second $\Z_2$ subgroups respectively and $v$ for the diagonal $\Z_2$ subgroup. 
Since $H^2(\Z_2 \times \Z_2,U(1))=\Z_2$, this is the simplest setup where an internal symmetry $G$ exhibits an SPT. This will be contrasted with how the SPT for time-reversal symmetry arises in the SymTFT in the next section. 

The SymTFT has sixteen topological lines, which we can label as $a_{g,ij}$ following the notation of the previous section. Here $g \in \{ 1,s,c,v \}$ represents a $G$ conjugacy class (which consists of a single element since the group is abelian) and by $ij$, with $i,j \in \{+,- \}$, we denote an irrep of $\Z_2 \times \Z_2$.  The symmetry boundary realising $\Z_2 \times \Z_2$ is given by 
\begin{align}\label{eq:Z2Z2_Dir}
    \cL_\Dir = a_{1,++} \oplus a_{1,+-} \oplus a_{1,-+} \oplus a_{1,--} \,.
\end{align}

The trivial SPT phase is obtained by choosing for the physical boundary 
\begin{align}
    \cL_\Neu =  a_{1,++} \oplus  a_{s,++} \oplus  a_{c,++} \oplus a_{v,++} \,.
\end{align}
Indeed, since only the identity line $a_{1,++}$ can end on both boundaries, we  get a single vacuum. The string order parameters correspond to the anyons appearing in $\cL_\Neu$. These are clearly uncharged under the symmetry generators $s,c,v$ of $\Z_2 \times \Z_2$, and therefore this is the trivial SPT phase. 

The non-trivial SPT phase is obtained by choosing for the physical boundary 
\begin{align}
     \cL_{\Neu,b} =  a_{1,++} \oplus  a_{s,+-} \oplus  a_{c,-+} \oplus a_{v,--} \,.
\end{align}
As above, only the identity line $a_{1,++}$ can end on both boundaries, and therefore we get a single vacuum. String order parameters correspond to anyons appearing in $\cL_{\Neu,b}$, and this time they are charged under the $\Z_2 \times \Z_2$ symmetry. In particular, we have an $s$-twisted sector order parameter charged $-1$ under $c,v$, a $c$-twisted sector order parameter charged $-1$ under $s,v$, and a $v$-twisted sector order parameter charged $-1$ under $s,c$. Notice this follows from the braiding of the anyons in the bulk, in particular the $-1$ braiding between $a_{s,+-}$ and $a_{c,++}$ and the $-1$ braiding between $a_{c,-+}$ and $a_{s,++}$(and similarly for the other cases). These charges are the hallmark of a non-trivial SPT phase\footnote{All the other Lagrangian algebras in $\cZ(\Z_2 \times \Z_2)$ share anyons other than $a_{1,++}$ with \eqref{eq:Z2Z2_Dir}, and therefore give rise to gapped phases with multiple vacua when selected as $\Bphys$. Systematically analyzing these cases recovers the expected symmetry-breaking phases \cite{Bhardwaj:2023idu}.}. 

Since we are in a symmetric phase with a single vacuum, all the line operators in $\Z_2 \times \Z_2$ are identified with the identity line operator. The difference compared to the trivial symmetric phase is the presence of non-trivial junction operators between lines. 
The non-trivial junctions are as follows 
\begin{align}
\includegraphics[]{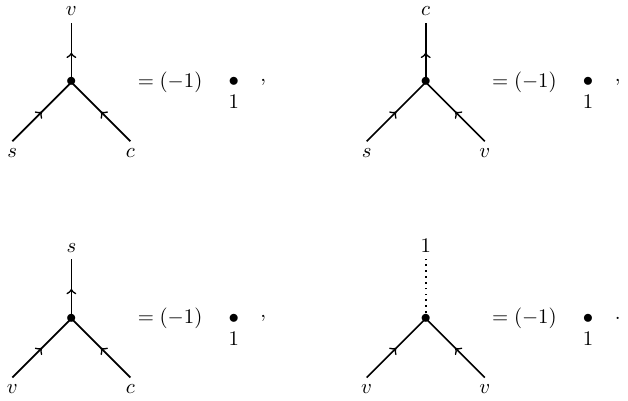}\label{eq:junctions}
\end{align}
As we will see, this will be important to recover the correct charge of the order parameters in the IR $(1+1)$d TQFT coming from the interval compactification of the SymTFT with physical boundary specified by $\cL_{\Neu,b}$.

For generic group $G$, we dress the junction operators between line operators by a 2-cochain $b(g,h)$ valued in $U(1)$. Associativity for the product of three lines $g,h,k$ implies that $b$ is actually a 2-cocycle. Moreover, this can be changed by an exact 2-cocycle by shifting the morphism between a line $g$ and the identity line by a 1-cochain $\gamma(g)$. Then $b \in H^2(G,U(1))$, and it turns out this is precisely the cohomology class specifying the discrete torsion label for the physical boundary $\cL_{\Neu(G),b}$. Pictorially, we have  
\begin{align}
\includegraphics[]{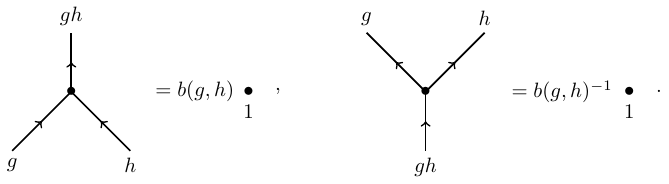}\label{eq:junctions2}
\end{align}

\subsection{String order parameters and selection rules in the SymTFT} \label{sec:stringorder_SymTFT}
In \Cref{sec:stringorder} we reviewed the arguments for string order parameters on the lattice. Here we outline how the same idea appears in the standard SymTFT setting \cite{Bhardwaj:2023idu}.

We assume that we are in an SPT phase with unique vacuum. A key point in this setting is that, unlike on the lattice, we cannot consider any non-local operator, but only those that survive in the low-energy limit. In fact, there is no meaning to choosing different end-point operators for our symmetry strings. In particular, in the IR our symmetry strings act trivially, and the only local operator is proportional to the identity.
This is captured in the following identity
\begin{align}
\includegraphics[]{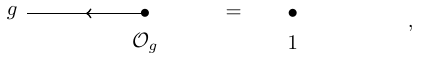}\label{eq:string1}
\end{align}
where $\mathcal{O}_g$ is a $g$-twisted sector local operator. 

This means that our lattice point of view must be slightly modified. By working in the IR we are automatically considering non-vanishing string order parameters. We then wish to find consistency relations relating the end-point charge to the SPT phase that is encoded in the junctions of lines in our category \eqref{eq:junctions2}. Another perspective is that we show string operators that survive in the IR necessarily have the `correct end-point charge' to balance out the symmetry string. This gives the same conclusion as on the lattice: if a UV string operator has the wrong end-point charge, it will not appear in the IR, and hence not have long-range order.
 
Let us fix the case where we have two commuting symmetries $g$ and $h$. We are looking to identify the discrete torsion phase as in \Cref{sec:discretetorsion}.
Let us consider the charge $e^{i\chi_{g,h}^{\mathcal{O}_g}}$ of the end-point, which we define as appearing in the following move:
\begin{align}
\includegraphics[]{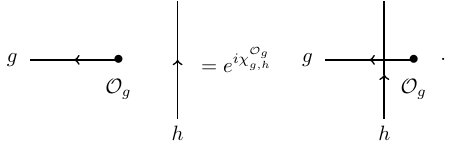}\label{eq:string2}
\end{align}
Notice that the charge $e^{i\chi^{\cO_g}_{g,h}}$ is a `prediction' of the SymTFT: a $g$-string order parameter comes from a bulk anyon $([g],\pi)$ ending on the physical boundary, and therefore $e^{i\chi^{\cO_g}_{g,h}} = \pi(h)$ (for $g$ and $h$ commuting).

We can resolve the right-hand-side diagram as follows
\begin{align}
\includegraphics[]{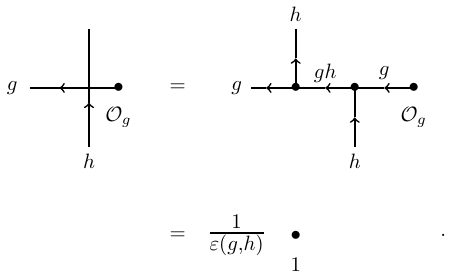}
\end{align}
The phase $\eps(g,h)^{-1}=b(h,g)/b(g,h) = \eps(h,g)$ comes from applying \cref{eq:junctions2}. The value is an invariant of the SPT phase of the vacuum. Now, the $h$ line on the left-hand-side of \cref{eq:string2} acts as the identity, so it can be freely inserted and removed, and we conclude that 
\begin{align}
\includegraphics[]{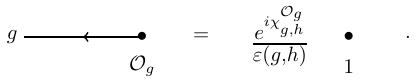}
\end{align}
Unless ${e^{i\chi_{g,h}^{\mathcal{O}_g}}}={\eps(g,h)}$, this is inconsistent with \cref{eq:string1}. We have hence correlated the analogue of the end-point charge with the analogue of the symmetry fractionalization (the diagrammatic rules for the fusion category). Compare to \cref{fig:unitarySO} where we analyzed a $g$-twist string operator and considered the charge under a commuting symmetry $h$.

Note that putting our TFT on a torus, analogous manipulations show us that $\eps(g,h)^{-1}=
\eps(h,g)$ is the appropriate $(g,h)$-twisted torus partition function.
\begin{align}
\includegraphics[]{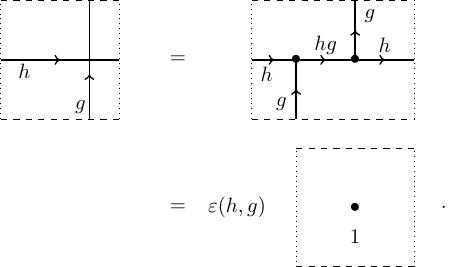}
\end{align}

\section{SymTFT with time-reversal symmetry} \label{sec:timereversal}

In this section, we consider a generalization of the SymTFT picture above in the case where the symmetry category $\cC$ contains time-reversing elements. In particular, we study a theory with symmetry involving an internal component, described by a fusion category $\cC_1$, and possibly some time-reversing elements.
This follows the proposal for space-time symmetry-enriched SymTFT \cite{Pace:2025hpb}.

\subsection{A graded category for time-reversal}

The first thing we need to do is to understand how to incorporate the anti-unitary nature of time-reversal in the categorical description. Our treatment is along the lines of  that given in \cite{cheng2017exactly,Bhardwaj:2016dtk,Barkeshli:2016mew}. 

Denote by $\cC$ the category of lines combining the internal symmetry and the time-reversing elements. This comes with a homomorphism 
\begin{align}
    \epsilon: \cC \rightarrow \Z_2 \,,
\end{align}
which specifies if an element in $\cC$ is time-reversing or not. The kernel of this homomorphism is clearly the internal symmetry, which we denote by $\cC_1$. The collection of lines $\cC  \setminus  \cC_1$ gives the time-reversing elements. In the simplest case where $\cC_1 = \Vec$ and $\cC= \Z_2^T = \{1,T \}$, $\epsilon$ assigns $\epsilon(T)=-1$ to the non-trivial element of $\Z_2^T$. This gives a $\Z_2$ grading to the category $\cC$. More precisely, to incorporate time-reversal we assign a plaquette label $i = \pm 1$ to the regions between two lines. The total symmetry category $\cC$ is then comprised of two sectors $\cC_1  \oplus \cC_{-1}$, where $\cC_{-1}$ indicates the possible labellings of lines that sit between a $+1$ plaquette and a $-1$ plaquette (and viceversa), while $\cC_1$ contains the internal symmetry lines. For $\cC$ to be a well-defined $\Z_2$ graded category, it is important that the fusion rules respect the grading structure. That is, fusion of two lines in $\cC_i$ and $\cC_j$ respectively must produce a line in $\cC_{ij}$, $i,j=\pm1$\footnote{For simplicity, in the figure below we are assuming all  the vector spaces Hom$(a \otimes b, c)$ are one dimensional, as this is the case in all the examples we will encounter in this paper. Otherwise, one has to introduce an additional label for the vector space at the junction.}.
\begin{align}
\includegraphics[]{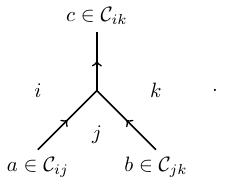}
\end{align}
Notice that the labelling of plaquettes is almost redundant. Once the leftmost plaquette, say, is fixed, all the other labellings can be inferred from the labels of the lines. 

The anti-unitary nature of time-reversal enters the category in the following important way. Compared to the standard unitary case, in our $\Z_2$ graded category $\cC$ the $F$-symbol acquires  a plaquette label $i$
\begin{align}
\includegraphics[]{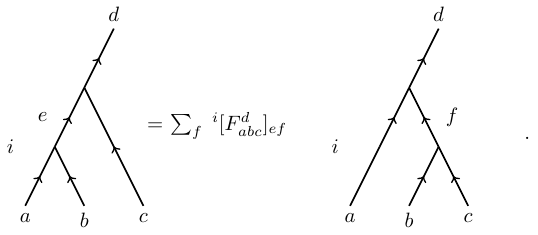}
\end{align}
In particular, $^{-1} [F]$ can be seen as the result of a time-reversal action on $^{+1} [F]$. Therefore, if we canonically define the $F$-symbol in the case where the left-most plaquette is labelled $+1$, the same $F$-symbol but with the left-most plaquette labelled $-1$ will be related by complex conjugation 
\begin{align}\label{eq:Fsymbol_twist}
    ^{-1} [F_{abc}^d]_{ef} = \left( ^{+1} [F_{abc}^d]_{ef} \right)^* \,.
\end{align}
Assuming that the left-most plaquette is labelled by +1, this leads to a `twisted' pentagon equation \cite{cheng2017exactly, Bhardwaj:2016dtk}
\begin{align}\label{eq:twist_pentagon}
    [F^e_{fcd}]_{gl} [F^e_{abl}]_{fk} = \sum_h [F_{abc}^g]_{fh} [F^e_{ahd}]_{gk} ([F^k_{bcd}]_{hl})^{\epsilon(a)} \ .
\end{align}
When the symmetry is a group $G$, the $F$-symbols reduce to a 3-cocycle $\omega(g,h,k)$, and the twisted pentagon equation is precisely the cocycle condition defining a class in $H^3_\epsilon(G,U(1))$. 

\subsection{Symmetry enriched SymTFT}\label{sec:SymEnriched}

Suppose our symmetry category is the graded fusion category $\cC$ introduced above. Following the proposal in \cite{Pace:2025hpb}, we consider as SymTFT for $\cC$ the double $\cZ(\cC_1)$  enriched by the $\Z_2^T$ anti-unitary symmetry. We denote this by $\cZ(\cC_1)^\times_{\Z_2^T}$. Let us comment on the relation between $\cZ(\cC_1)^\times_{\Z_2^T}$ and $\cZ(\cC)$. 

In general, consider a $G$-graded fusion category $\cC_G =\oplus_{g\in G} \cC_g$, with $G$ an internal unitary symmetry. One can use the standard Drinfeld center construction and obtain $\cZ(\cC_G)$. This contains a subcategory of anyons labeled by Rep$(G)$, which can be condensed to obtain $\cZ(\cC_1)^\times_G$. Mathematically, this represents a $G$-crossed braided extension of $\cZ(\cC_1)$, while physically it is an enrichment by the symmetry $G$ of the topological order $\cZ(\cC_1)$  \cite{Barkeshli:2014cna}. $\cZ(\cC_1)^\times_G$ is itself graded by $G$,  $\cZ(\cC_1)^\times_G = \bigoplus_{g\in G} \cZ(\cC_1)_g$. Here $ \cZ(\cC_1)_1 \equiv \cZ(\cC_1) $ describes the untwisted anyons in the Drinfeld center of $\cC_1$, while $\cZ(\cC_1)_g$ contains the $g$-defects (topological lines at the end of a $g$-surface). Fusion rules are compatible with the grading, i.e.\ a $g$-defect and an $h$-defect fuse into a $gh$-defect. Gauging $G$ in $\cZ(\cC_1)^\times_G$ then recovers $\cZ(\cC_G)$; mathematically, this process is known as $G$-equivariantization. For more details, see e.g.\ \cite{EGNO,Barkeshli:2014cna}. In summary, 
\begin{align}
\includegraphics[]{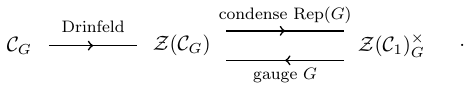}
\end{align}
The important point is that in $\cZ(\cC_G)$, i.e.\ in the usual SymTFT for $\cC_G$, the symmetry $G$ is gauged\footnote{Mathematically, there is a 1-to-1 correspondence between equivalence classes of $G$-extensions of $\cC_1$ and  $G$-crossed braided extensions of  $\cZ(\cC_1)$ \cite{etingof2010fusion}, at least in the unitary case.}.

What does this imply when the enriching symmetry is a space-time symmetry? An  appropriate background gauge field for time-reversal is given by the first Stiefel-Whitney class associated to the manifold $M_2$ where the theory is defined, which we denote by $w_1 \in H^1(M_2,\Z_2)$ \cite{Kapustin14}. This represents the obstruction to defining a consistent orientation on the manifold, i.e.\ a non vanishing $w_1$ implies that the manifold is unorientable. The orientation flips if we cross a locus homologous to the Poincaré dual $l_1 \in H_1(M_2,\Z_2)$ of $w_1$, which therefore can be understood as a place where the codimension-1 defect generating $\Z_2^T$ is inserted. 
Gauging a symmetry requires summing over background gauge fields in the path integral, and since the background gauge field for time-reversal concerns the nature of the space-time manifold, this takes us towards quantum gravity. The latter would naturally include a sum over space-time manifolds, and gauging time-reversal would involve summing over both orientable and non-orientable manifolds \cite{Harlow:2023hjb}.
Since we are concerned with quantum field theory and we do not have a clear prescription to gauge time-reversal, we  keep $\Z_2^T$ as a background symmetry. That is, we  only consider the SymTFT for the unitary symmetry $\cZ(\cC_1)$ enriched by time-reversal, rather than the full $\cZ(\cC)$.  The question we will explore in the following is precisely how much of the information in $\cZ(\cC)$ we can expect to recover by considering $\cZ(\cC_1)^\times_{\Z_2^T}$. 

We now recall the classification of the ways we can enrich $\cZ(\cC_1)$ by a symmetry $G$. This is spelled out in \cite{Barkeshli:2014cna}, at least for $G$ unitary (see \cite{Barkeshli:2016mew,Barkeshli:2017rzd,cheng2017exactly,heinrich2016symmetry} for extensions of this framework to anti-unitary symmetries.).
 For our purposes, we restrict to symmetry enriched topological phases where the underlying topological order is a Drinfeld center $\cZ(\cC_1)$. Again, we abstract one moment from our specific $\Z_2^T$ case of interest, and consider an enrichment by a generic group $G$: $\cZ(\cC_1)^\times_G$\footnote{Notice that our discussion of an appropriate time-reversal category changes only mildly when we consider extending $\cC_1$ by a group $G$ which may contain anti-unitary elements. In this case, $G$ itself would be graded to specify which elements are time-reversing.}. 
The classification is in terms of a triple $(\rho,\omega,\alpha)$, where
\begin{itemize}
    \item[1.] $\rho: G\rightarrow \text{Aut}(\cZ(\cC_1))$ specifies how the symmetry acts by permutation on the anyons;
    \item[2.] $\omega \in H^2_\rho(G,\cA)$ is the symmetry fractionalization class. Here $\cA$ denotes the subset of abelian anyons in $\cZ(\cC_1)$. This can be defined consistently when a certain obstruction class in $H^3_{\rho}(G,\cA)$ is trivial;
    \item[3.] $\alpha \in H^3(G,U(1))$ is the defect class, which corresponds to stacking a $G$-SPT phase and requires an obstruction in $H^4(G,U(1))$ to vanish.
\end{itemize}
We also recall the following useful fact. Suppose that, under some process, we have a phase factor $e^{i\phi_a} $ associated to each anyon $a$, compatible with the $\cZ(\cC_1)$ fusion rules:  $e^{i\phi_a} e^{i\phi_b} = e^{i\phi_c}$ if $N_{ab}^c \neq 0$. Then, there exists an abelian anyon $t(a) \in \cA$ such that the phase $e^{i\phi_a}$ obtained by $a$ under such process can be expressed as the braiding $M_{a,t(a)}$ of $a$ with $t(a)$ \cite{Barkeshli:2014cna}.

\subsection{Gapped boundaries of the SymTFT}

To study the gapped boundaries of the symmetry enriched SymTFT, we follow the mathematical treatment in Ref.~\cite{bischoff2019spontaneous} (and the related one in \cite{meir2012module}, where the authors classify module categories over graded fusion categories), as well as the more physical approach  in \cite{Cheng:2020rpl,Pace:2025hpb}. We should note that here we focus on $G$ unitary; we describe the novelties that appear when the enriching symmetry is anti-unitary in the examples. The question we are interested in is when a gapped boundary of $\cZ(\cC_1)$, specified by a Lagrangian algebra $\cL \in \cZ(\cC_1)$, preserves the enriching symmetry $G$. 

First of all, $\cL$ should clearly be invariant under the $G$ action, meaning that $\rho_g(\cL) = \bigoplus_{a} n_a \, \rho_g(a)\cong\cL$  for all $g \in G$. If a Lagrangian algebra $\cL$ does not satisfy this, we say that the boundary breaks the $G$ symmetry explicitly.
 
Second, the fractionalization class $w \in H^2(G,\cA_\cL)$ should be trivial, where $\cA_\cL$ denotes the set of uncondensed abelian anyons on the boundary. Equivalently, there should be no projective phases $\eta_a \in H^2(G,U(1))$ for any of the anyons $ a \in \cL$\footnote{Here we are assuming for simplicity that $\rho = \mathbbm{1}$, as will be the case in all the examples considered in this paper. The discussion can be generalized to non-trivial permutation action on the anyons.}.  This follows from the fact that $\eta_a (g_1,g_2) = M_{a,w(g_1,g_2)}$ and $\cL$ being maximal, in the sense that each anyon not in $\cL$ braids non-trivially at least with one anyon in $\cL$. When this condition is not satisfied, the symmetry $G$ is  spontaneously broken by the boundary specified by $\cL$ \cite{Jiang17}.
This condition implies also that the symmetry realized on the boundary is $\cA_\cL \times G$ (or possibly $\cA_\cL \rtimes G$ if the action $\rho$ is non-trivial), rather than an extension of $G$ by $\cA_\cL$, as would happen if $w(g_1,g_2) =b$ for some $b \in \cA_\cL$.  
Once this obstruction vanishes, we can have multiple solutions, parametrized by $t \in H^1(G,\cA_\cL)$. Mathematically, this corresponds to the different $G$-equivariant structures that we can put on the algebra $\cL$ \cite{bischoff2019spontaneous}. Physically, this can be interpreted  as assigning a 1d charge $\lambda_a(g)$ to each of the condensed anyons $a$ under the enriching symmetry $G$ (which can be expressed as the braiding with an abelian anyon modulo the condensed anyons, i.e.\ $\lambda_a(g) = M_{a,t(g)}$ with $t \in H^1(G,\cA_\cL)$) \cite{Jiang17,Cheng:2020rpl}. 
    
\begin{figure}
    \centering
  \includegraphics[]{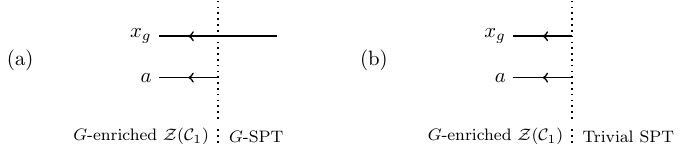}
    \caption{(a) A $G$-symmetric gapped boundary (dotted) of the enriched topological order $\cZ(\cC_1)$ provides in general an interface to a $G$-SPT. (b) If the SPT on the right hand side is the trivial SPT, then there is no obstruction to terminate the $G$-twists on the boundary.}
    \label{fig:defects_end}
\end{figure}

Suppose we have an $\cL$ which satisfies the above conditions. Then this naturally defines a gapped interface between $\cZ(\cC_1)^\times_G$ and an invertible theory with $G$ symmetry, which is a $G$-SPT, classified by $H^3(G,U(1))$. In this case, the $g$-defects of $\cZ(\cC_1)^\times_G$ `pass through' the boundary and become the defects of the $G$-SPT. However, if the boundary is fully $G$ symmetric and the phase on the other side is the trivial $G$-SPT, so that the boundary is genuine rather than an interface with a non-trivial SPT, then also the $g$-defects condense at the boundary. See \cref{fig:defects_end}. 
Consistency with the bulk fusion rule of two twist defects requires the existence of a boundary $M$-symbol for the $g$-defects
\cite{Cheng:2020rpl}:
\begin{align}\label{eq:twist_end}
\includegraphics[]{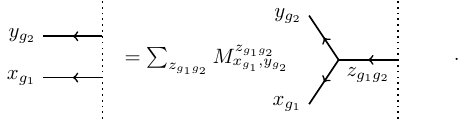} 
\end{align}
These $M$-symbols are subject to a pentagon equation, involving four $M$-symbols and one $F$-symbol for the $G$-defects. Schematically,\footnote{This equation will be twisted by $\epsilon$ when the enriching symmetry involves anti-unitary elements, as we discuss in \cref{sec:Z2T_phases}.}
\begin{align}\label{eq:Msol_general}
    \sum_{w_{g_1g_2}} M_{x_{g_1},y_{g_2}}^{w_{g_1g_2}} M_{w_{g_1g_2},z_{g_3}}^{k_{g_1g_2g_3}} (F_{x_{g_1}y_{g_2}z_{g_3}}^{k_{g_1g_2g_3}})_{w_{g_1g_2} h_{g_2g_3}} = M_{y_{g_2} ,z_{g_3}}^{h_{g_2g_3}} M_{x_{g_1}, h_{g_2 g_3}}^{k_{g_1 g_2 g_3}} \,.
\end{align}

We can then also interpret the 1d charge $\lambda_a(g)$ of the condensed anyon $a$ under $G$ as the following move 
\begin{align}
\includegraphics[]{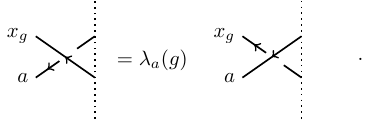} 
\end{align}
Considering crossing by two $G$-defects $x_{g_1}$ and $x_{g_2}$, we obtain the consistency equation 
\begin{align}
    \eta_a(g_1,g_2) = \frac{\lambda_a(g_1g_2)}{\lambda_a(g_1)\lambda_a(g_2)} \,.
\end{align}
Therefore we again find the condition that the fractionalization class should be trivial.

We also recall that if the action $\rho$ is trivial as we are assuming, then there exists at least one abelian $g$-defect $1_g$ for every $g\in G$, such that the other $g$-twists can be obtained as $a_g = a \otimes 1_g$, with $a \in \cZ(\cC_1)$ \cite{Barkeshli:2014cna}. Their fusion rule in general reads
\begin{align}
    a_{g_1} \otimes b_{g_2} = [a \otimes b \otimes w(g_1,g_2)]_{g_1g_2} \,.
\end{align}

We can always choose what we call a canonical boundary condition, where all the anyons $a \in \cL$ are uncharged under the enriching symmetry. This corresponds to a canonical choice of $g$-defects on the other side of the gapped boundary given by $1_g$. When instead a condensed anyon $a$ is charged, this amounts to a dressing of the $g$-defect, which becomes $1_g \rightarrow 1_g \otimes t(g)$. The new dressed defects might have non-trivial $F$-symbols, which effectively means there is an obstruction to the solution of \cref{eq:Msol_general}. This implies the twist defects have to continue into an SPT  characterized by $H^3(G,U(1))$. Therefore, not all charge assignments are compatible with  ending the $g$-defects on the gapped boundary. We will see this explicitly in examples below, and argue that when this obstruction vanishes we obtain a solution valued in $H^2(G,U(1))$ (recall, moreover, that $w(g_1,g_2)$ has to be condensed on the boundary from the second condition above).

In this way, the discussion of $G$-symmetric gapped boundaries of $\cZ(\cC_1)^\times_G$ can be framed in terms of a series of obstructions and solutions to those obstructions. This is in the same spirit as the classification of $G$-enriched SETs, see also \cite{meir2012module}.

\section{Examples of classification of phases via SymTFT}\label{sec:examples}

In this section we apply the classification of gapped boundaries outlined above to determine gapped phases with time-reversal symmetry. Before doing so, we will warm up with an internal symmetry example. One important comment is that since we restrict our attention to boundaries that preserve the enriching symmetry, we will classify only those gapped phases that do not (spontaneously) break the enriching symmetry. This is intimately related to our inability to gauge the enriching symmetry (at least in the anti-unitary case). 

\subsection{$\Z_4$ gapped phases}\label{sec:Z4_unitary}

We think of $\Z_4$ as an extension of $\Z_2^s$ by $\Z_2^m$
\begin{align}
    1 \rightarrow \Z_2^m \rightarrow \Z_4 \rightarrow \Z_2^s \rightarrow 1 \,.
\end{align}
This can be seen as a $\Z_2^s$ graded category where the non-trivial element in $\Z_2^s$ squares to the non-trivial element in $\Z_2^m$.
The corresponding SymTFT is the toric code $\cZ(\Z_2^m)$ enriched by $\Z_2^s$. The aim is to classify the phases that do not spontaneously break the enriching symmetry $\Z_2^s$. The anyons in the toric code are $\{ 1,e,m,f\}$, and in addition we have the $s$-twisted sectors $\{ 1_s,e_s,m_s,f_s\}$\footnote{For the untwisted anyons, we leave the subscript $1$ implicit.}. The fact that the input symmetry is $\Z_4$ implies a non-trivial fractionalization class for the SymTFT, i.e.\ we have $w(s,s)=m$ and therefore the $e$ anyon carries a projective phase $\eta_e(s,s)=-1$. The symmetry action $\rho$ is trivial, i.e. there is no permutation of the anyons.

Now let us examine the boundary conditions. We have $\cL_e = 1 \oplus e$, which would be the natural choice as we want to realize the $\Z_2^m$ symmetry on the boundary. However, this choice is not allowed: although  $\rho(\cL_e)=\cL_e$, $e$ carries a projective $\Z_2^s$ charge and this implies that the gapped boundary spontaneously breaks the enriching symmetry. The other boundary condition of the toric code is $\cL_m$. This is $\Z_2^s$ symmetric as $\rho(\cL_m)=\cL_m$ and $\eta_m(s,s)=1$. Notice, however, that this choice realizes $(\Z_2^e \times \Z_2^s)^\nu$ on the boundary, rather than $\Z_4$. The latter is related to the former via the gauging of $\Z_2^m$, and the anomaly $\nu$ is a consequence of the symmetry extension \cite{Tachikawa:2017gyf}. We thus persist to see what we can learn about the $\Z_4$ case.

Consider first the canonical boundary condition for $\cZ(\Z_2^m)^\times_{\Z_2^s}$, where $m$ carries no charge under $\Z_2^s$. This implies that we also end the $s$-defect $1_s$. The solution for the $M$-symbol in \cref{eq:Msol_general} reduces to a class in $H^2(\Z_2^s,U(1))$, which is trivial. Therefore, we get only one gapped boundary. Notice that $(1_s)^2 = m$. So, when we solve for the boundary $M$-symbol, in principle this would depend on both $1_s$ and $m$. However, we are solving for this on the $\cL_m$ boundary, where $m$ is being condensed: $m \sim 1$. Therefore, the non-trivial $M$-symbol---on this specific boundary--- effectively depends only on $1_s$, i.e.\ it is a group 2-cocycle $b(1_s,1_s)$. This is why the solution to \cref{eq:Msol_general} gives an $H^2$ class for the enriching symmetry\footnote{In this analysis, we assume we can work in a `step-wise' manner and solve first for a boundary defined via condensation of the untwisted anyons, on which we then end the twist defects.}.

Now consider the possibility of having the $m$ anyon charged under the enriching symmetry $\Z_2^s$. This amounts to a shift in the $s$-defect: $1_s \rightarrow e \otimes 1_s = e_s$. The $-1$ charge of $m$ comes from the braiding with the $e$ anyon that we have dressed the $s$-defect with. An important difference with the previous case is that, due to the fractional charge $\eta_e(s,s)=-1$, $e_s$ is a semion. In particular, it has an anomaly $\omega(s,s,s) = M_{w(s,s),t(s)} = M_{m,e}=-1$, and, when continuing to the other side of the $\cL_m$ interface,  becomes the twist defect for the non-trivial $\Z_2^s$ SPT phase specified by the non-trivial cocycle in $H^3(\Z_2^s,U(1))$ \cite{bischoff2019spontaneous,Jiang17}. This implies also that the solution to \cref{eq:Msol_general} is obstructed in this case by the non-vanishing anomaly. We learn that the boundary where the $m$ anyon carries $\Z_2^s$ charge can be an interface only to a non-trivial SPT phase, and cannot be realized as a boundary to the vacuum. Therefore, we discard this possibility in our study of (1+1)d gapped phases. 

In summary, we have only one $\Z_2^s$ symmetric boundary, $\cL_m$ with $m$ uncharged under $\Z_2^s$. Selecting this as both the symmetry and the physical boundary, we obtain a gapped phase with two vacua that spontaneously breaks the $\Z_2^e$ symmetry---this comes from the anyon $m$ ending on both boundaries, giving rise to a local operator $\cO_m$ charged under $\Z_2^e$. This is consistent with our expectation that the enriched SymTFT would capture only phases that do not spontaneously break the enriching symmetry (the other phases would be $\Z_2^s$ SSB and full $(\Z_2^e \times \Z_2^s)^\nu$ SSB). 

We can map this to a gapped phase for $\Z_4$ by gauging the internal symmetry $\Z_2^e$. This gets mapped to the $\Z_4$ trivial phase, again the only phase for $\Z_4$ that does not spontaneously break $\Z_2^s$.

\subsection{$\Z_2^T$ gapped phases}\label{sec:Z2T_phases}

In this case, the input symmetry category is the anti-unitary  $\Z_2^T$ graded category $\cC$ with trivial $\cC_1 = \Vec$. Correspondingly, our SymTFT is the trivial topological order enriched by $\Z_2^T$. This contains only the trivial line $\{ 1\}$, and the $T$-twisted sector line $\{ 1_T\}$. The only Lagrangian algebra is $\cL = 1$, which is clearly $\Z_2^T$ symmetric. 

We should then provide a solution for \cref{eq:Msol_general}.  In this case, the only non-trivial $M$-symbol dependence is on the $T$-defect, i.e.\ we simplify to $M = b(1_T,1_T)$. Moreover, similar to the discussion around \cref{eq:Fsymbol_twist}, due to the anti-unitary nature of the symmetry, the $M$-symbol will be conjugated if applied on an initial plaquette labelled by $-1$. 

\begin{figure}
\includegraphics[]{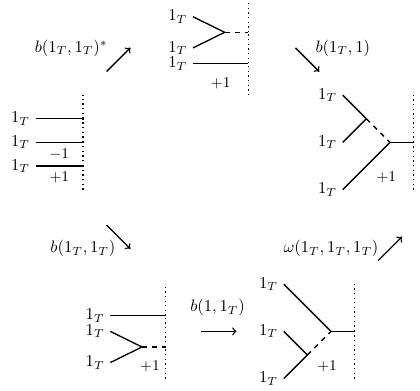}
\caption{The pentagon equation satisfied by $M=b(1_T,1_T)$.}\label{fig:pentagon}
\end{figure}

Consider then a configuration with three $1_T$ lines ending on the boundary, as in \cref{fig:pentagon} (the identity line is indicated by a dashed line).  Without loss of generality, we start again with a configuration where the initial plaquette is labelled by $+1$. As shown below, all the $M$-moves involve a $+1$ plaquette, except for the first move on the upper path, which involves starting with a plaquette labelled $-1$\footnote{In \cref{fig:pentagon} we omit the arrows since the symmetry is $\Z_2$.}.

We work in the usual gauge where $b(\cdot,1)= b(1,\cdot)=1$ and we assume the $\Z_2^T$ symmetry is non-anomalous, i.e.\ $\omega(1_T,1_T,1_T) = 1$. Therefore we get 
\begin{align}
    b(1_T,1_T) = \pm 1 \,,
\end{align}
i.e.\ we recover the non-trivial cocycle $H^2_\epsilon(\Z_2^T,U(1)) = \Z_2$.

We can easily generalize this to $\mathcal{C}_1$ extended by a generic group $G$, which can possibly contain anti-unitary elements. Consider three lines $1_{g_1}$, $1_{g_2}$, $1_{g_3}$ ending on the boundary. Performing the same moves as in \cref{fig:pentagon}, $1_{g_2}$ is the first line  for which we apply the $M$-move with a left-most plaquette labelled by $+1$ or $-1$ depending on whether $g_1$ is time reversing or not. Therefore we find 
\begin{align}
    b(1_{g_1},1_{g_2}) b(1_{g_1 g_2}, 1_{g_3}) =\; b(1_{g_2},1_{g_3})^{\epsilon(g_1)} b(1_{g_1},1_{g_2g_3}) \,, 
\end{align}
i.e.\ $b$ is now a cocycle in $H^2_\epsilon(G,U(1))$. 

Coming back to the SymTFT sandwich for $G=\mathbb{Z}_2^T$, we see that from $\cL=1$ we obtain two boundary conditions, $\cL_\pm$ depending on $b(1_T,1_T)$. Let us fix the symmetry boundary to be $\cL_+$. Choosing $\cL_+$ as the physical boundary gives the trivial phase. Choosing $\cL_-$ as the physical boundary gives the non-trivial $\Z_2^T$ phase. In this latter case, we decorate the junction of two $T$-lines by a sign in the resulting $(1+1)$d gapped phase.

\subsection{$\Z_2 \times \Z_2^T$ gapped phases}

The symmetry category here is a trivial extension of $\Z_2^T$ by $\Z_2^m$, i.e.\ two $T$ lines square to the identity. The corresponding SymTFT is the toric code $\cZ(\Z_2^m)$ enriched by $\Z_2^T$. We denote the $T$-defects by $\{1_T, e_T,m_T,f_T \}$. The symmetry enrichment is trivial, in the sense that $\rho = \mathbbm{1}$ and also the symmetry fractionalization is $w(1_T,1_T)=1$. This means, in particular, that any gapped boundary of the toric code is automatically $\Z_2^T$ symmetric. 

Let us first consider $\cL_e = 1 \oplus e$. We have the canonical boundary, where $e$ carries no charge under $\Z_2^T$. This amounts to ending $1_T$ on the boundary. From the discussion in the previous section, we know that this has two different solutions classified by $H^2_\epsilon(\Z_2^T,U(1)) = \pm1$. We can also consider the boundary where $e$ carries $-1$ charge under $\Z_2^T$. This corresponds to ending $m_T$ on the boundary. This is equivalent to ending $1_T$, and therefore we also get two possible solutions. In summary, we obtain four $\Z_2^T$-symmetric gapped boundaries, which we denote by $\cL_{e,+}$, $\cL'_{e,+}$, $\cL_{e,-}$, $\cL'_{e,-}$, where $\cL'_e$ denotes the boundary where $e$ is charged and $\pm $ denotes the $\Z_2^T$ SPT. The discussion is completely analogous for $\cL_m$, and we get four boundaries $\cL_{m,+}$, $\cL'_{m,+}$, $\cL_{m,-}$, $\cL'_{m,-}$. 

For the SymTFT sandwich, we fix the symmetry boundary
\begin{align}
    \Bsym = \cL_{e,+} \,,
\end{align}
which realizes $\Z_2^m \times \Z_2^T$. We then obtain eight gapped phases with that symmetry:
\begin{itemize}
    \item[1.] $\Bphys = \cL_{e,+}$: this is the usual $\Z_2^m$ SSB phase, with local order parameter $\cO_e$ coming from the anyon $e$ ending on both boundaries;
    \item[2.] $\Bphys = \cL_{m,+}$: this is the usual trivial phase with a single symmetric ground state (only  the identity line ends on both boundaries);
    \item[3.] $\Bphys = \cL'_{e,+}$: this is a $\Z_2$ SSB phase that preserves the diagonal $\Z_2$ symmetry of $\Z_2^m \times \Z_2^T$, as we obtain one local order parameter $\cO_e$ that is charged $-1$ under $\Z_2^m$ and $\Z_2^T$ and $+1$ under the diagonal $\Z_2$;
    \item[4.] $\Bphys = \cL'_{m,+}$: this is a $\Z_2^m \times \Z_2^T$ SPT phase, where the order parameter is an $m$-twist operator $\cO_m$ charged $-1$ under $\Z_2^T$. This comes from the anyon $m$ ending on the physical boundary. In terms of background gauge fields $a \in H^2(M_2,\Z_2^m)$ and $w_1 \in H^2(M_2,\Z_2^T)$, this would read $\pi i \int_{M_2} a \cup w_1$; 
    \item[5.] $\Bphys = \cL_{m,-}$: this is a $\Z_2^T$ SPT phase with a single symmetric ground state. The corresponding action is $\pi i \int_{M_2} w_1^2 $. There are no operators that can be interpreted as a string order parameter;
    \item[6.] $\Bphys = \cL_{e,-}$: this is the $\Z_2^m$ SSB phase stacked with the $\Z_2^T$ SPT;
    \item[7.] $\Bphys = \cL'_{e,-}$: this is the other `diagonal' $\Z_2^m$ SSB phase stacked with the $\Z_2^T$ SPT;
    \item[8.] $\Bphys = \cL'_{m,-}$: this is the stacking of the $\Z_2^m \times \Z_2^T$ SPT and the $\Z_2^T$ SPT.  
\end{itemize}
This reproduces all the expected $\Z_2^m \times \Z_2^T$ phases that do not spontaneously break the enriching $\Z_2^T$ symmetry, see e.g. \cite{Spieler:2025hyr}, and also the SPT cohomology classification $H^2_\epsilon(\Z_2\times \Z_2^T,U(1)) = \Z_2 \times \Z_2$ \cite{Chen13}.

\subsubsection{String order parameters for $\Z_2 \times \Z_2^T$:} Let us comment further on the string order parameter that we obtain in the mixed SPT phase (phase 4 in the above list). This is a unitary $m$-string with an endpoint $\cO_m$ charged $-1$ under the time-reversal symmetry
\begin{align}
\includegraphics[]{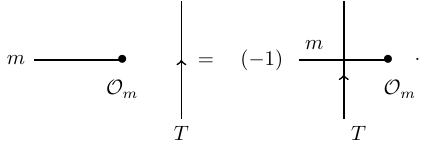}\label{eq:string3}
\end{align}
Analogous to the unitary symmetry case explained in \cref{sec:stringorder_SymTFT}, in the IR TQFT all the symmetry lines are identified with the identity, and any non-triviality is encoded in the choice of the junction operators between symmetry lines. This allows us to relate the $-1$ charge prescribed by the SymTFT to a gauge invariant combination of the $b(g,h)$. In this $\mathbb{Z}_2$ case we can fix a gauge where $b(m,m)=1$, which means that the complex-conjugation action in the above diagram cannot be gauged away. In particular, multiplying $\mathcal{O}_m$ by $i$ would toggle the $T$-charge, but would also change $b(m,m)$.
We then obtain $e^{i\chi^{\cO_m}_{m,T}} = (-1) = \kappa(T,m)$. This agrees with the MPS derivation in \cref{sec:stringorder} for $\Z_2$ unitary symmetry (for which we can always choose an hermitian endpoint operator). Note this is a standard `charged end-point' string order, rather than one of the exotic lattice order parameters
with two copies of the state, discussed in Refs.~\cite{Pollmann12,Shiozaki16}.

We also remark that for the $\Z_2^T$ SPT (phase 5 in the above list) we give another exotic MPS based string-order in \Cref{app:anti-unitarystring}. This is interpreted as the $T$-charge of the end-point of a $T$-string, but is only well-defined given the tensor network ground state. As one would expect, this does not appear within the enriched SymTFT framework.

\subsubsection{Comments on generalization to $\Z_n \rtimes \Z_2^T$:} We make some general remarks on the case where the unitary symmetry is $\Z_n$ rather than $\Z_2$. Notice the semi-direct product implies the fusion rule $TgT^{-1} = g^{-1}$ for every $g \in \Z_n$. This is consistent with the expectation that crossing a time-reversal defect $T$ switches the orientation of the line. 

Now let us consider the corresponding SymTFT. The input category being $\Z_n \rtimes \Z_2^T$ results in an enrichment of $\cZ(\Z_n)$ by $\Z_2^T$ with non-trivial action $\rho$. In particular, denoting a flux anyon of $\cZ(\Z_n)$ by $m^j$, $j=0,\dots,n-1$, we have $\rho(m^j)=m^{-j}$. 
Now imagine having the charged condensation of a pure flux $m^j$, potentially leading to a string order parameter for a mixed $\Z_n \rtimes \Z_2^T$ SPT. Due to the non-trivial $\rho$ action, we clearly see that this is possible only when $m^j = m^{-j}$, i.e.\ when $m^j$ is the generator of the $\Z_2$ subgroup of $\Z_n$. This also implies $n$ has to be even to have an SPT order parameter of this type. Only in this case can we assign a well-defined 1d charge (which is $\pm 1$). We interpret this as the endpoint charge of the $m^{n/2}$ string under $\Z_2^T$. 

This is consistent with the cohomology classification $H^2_\epsilon(\Z_n \rtimes \Z_2^T,U(1)) = \Z_2 \times \Z_{(2,n)}$ \cite{Chen11}, where the first $\Z_2$ is generated by the pure $\Z_2^T$ SPT and the second $\Z_{(2,n)}$ by the mixed SPT. Note that this $\Z_2$ subgroup of $\Z_n$ is clearly always preserved whenever we preserve $\Z_n$, so fixing the larger symmetry group does not add any further classes of projective representations. 

More generally, we can always detect an SPT phase by the value of the partition function on some particular manifold with insertion of topological lines \cite{Pollmann12}. In the presence of a unitary symmetry $G$ and anti-unitary $\Z_2^T$, the appropriate manifold to diagnose the SPT is the Klein bottle, which indeed has one oriented and one orientation-reversing 1-cycle.  The $(T,g)$-twisted partition function gives the invariant combination $\kappa(T,g) = \frac{b(T,g^{-1}) b(g,g^{-1})}{b(g,T)}$ appearing also in the MPS analysis in \cref{sec:stringorder}:
\begin{align}
\includegraphics[]{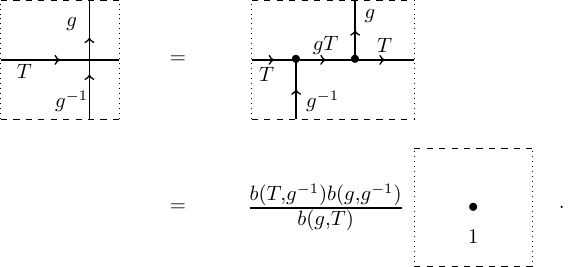}
\end{align}
Since a $g$-loop becomes $g^{-1}$ after crossing the orientation-reversing cycle, the consistency of the network requires the extra factor $b(g,g^{-1})$ coming from the junction where $g$ and $g^{-1}$ annihilate.

Notice that there are three concepts that in the case of a unitary symmetry $G$ with commuting elements $gh=hg$ we tend to conflate, as they are all equivalent and provide a good diagnostic for an SPT phase. These are (i) the torus partition function with $h,g$ symmetry line insertions, (ii) the $h$-charge of the $g$-twisted ground state, and (iii) the $h$-charge of the endpoint $\cO_g$ of a $g$-string. The latter two are related via the state-operator correspondence. The anti-unitary case is more subtle due to the orientation-reversing nature of $\Z_2^T$. Therefore, while the Klein bottle invariant is always well-defined, we cannot always interpret this straightforwardly as the charge of a unitary symmetry string endpoint. Indeed, this manifested in our discussion of end-point charges on the lattice in \cref{sec:klein}.

\subsection{$\Z_4^T$ gapped phases}

To realize $\Z_4^T$, the symmetry category is a $\Z_2^T$ graded fusion category  corresponding to the non-trivial extension
\begin{align}\label{eq:Z4T_ext}
    1 \rightarrow \Z_2^m \rightarrow \Z_4^T \rightarrow \Z_2^T \rightarrow 1 \,.
\end{align}
Therefore, the generator $T$ is anti-unitary, and it squares to the unitary element $m$, i.e.\ $T^2 = m$.

The corresponding SymTFT is again the toric code $\cZ(\Z_2^m)$ enriched by $\Z_2^T$, but this time with non-trivial symmetry fractionalization class. In particular, we have $w(T,T) = m$, which implies $\eta_e(T,T)=-1$, i.e.\ $e$ carries fractionalized charge under the $\Z_2^T$ symmetry. Compared to the previous $\Z_2 \times \Z_2^T$ case, this implies the $\cL_e = 1\oplus e$ boundary is not allowed. The only $\Z_2^T$ symmetric boundary is $\cL_m = 1\oplus m$.

Consider first the canonical boundary, where $m$ carries no charge under $\Z_2^T$. On top of it, we can also end $1_T$, and this gives us a class in $H^2_\epsilon(\Z_2^T,U(1)) = \Z_2$. We denote the two boundaries by $\cL_{m,\pm}$.

Now, consider the possibility of $m$ carrying a $-1$ $\Z_2^T$ charge. This amounts to shifting $1_T \rightarrow e \times 1_T = e_T$. 
Remember we excluded this case for a  $\Z_4$ unitary symmetry in \cref{sec:Z4_unitary}, as a non-vanishing obstruction in $H^3(\Z_2^s,U(1))$ led to not being able to end $e_s$ on the boundary.  Here we have however an important difference. For an anti-unitary $\Z_2^T$ symmetry, $H^3_\epsilon(\Z_2^T,U(1)) = \Z_1$, so we know any $\Z_2^T$ symmetric condensation necessarily leads to the trivial SPT on the right-hand-side of the boundary\footnote{One can check $\omega(T,T,T) =1$ is still compatible with $M_{w(T,T),t(T)} = M_{m,e} = -1$ thanks to a twisted hexagon identity that takes into account complex conjugation in appropriate places, analogously to \cref{eq:twist_pentagon} for the pentagon identity. In particular, for a unitary $\Z_2$ symmetry, we have from the hexagon identity the relation $(R_{ss}^1)^2 = \omega(s,s,s)=\pm1$, where $R_{ss}^1$ is the braiding of $s$ with itself. In the anti-unitary $\Z_2^T$ case, this becomes $R_{TT}^1 (R_{TT}^1)^* = \omega(T,T,T) = 1$. }. Therefore, there is no obstruction and we get also two boundaries $\cL_{m,\pm}'$, where the $m$ anyon being condensed is charged under $\Z_2^T$.

Moving to the sandwich, we select as symmetry boundary
\begin{align}
    \Bsym = \cL_{m,+} \,,
\end{align}
realizing $(\Z_2^e \times \Z_2^T)^\nu$.  We obtain the following gapped phases:
\begin{itemize}
    \item[1.] $\Bphys = \cL_{m,\pm}$: this is a $\Z_2^e$ SSB phase. However, we observe something new here. From the anyon $m$, we obtain a local operator $\cO_m$ charged under $\Z_2^e$ in the (1+1)d TQFT. From this and the identity operator, let us construct the two vacua $v_\pm = \frac{1\pm\cO_m}{2}$. These two are permuted by the broken $\Z_2^e$ symmetry. In particular, $v_-$ can be obtained by applying $e$ to $v_+$, i.e. $v_- = e (v_+)$. Recall however that $e$ carries fractional charge under $T$, i.e.\ $T^2(e) = -1$. This implies that the second vacuum $v_-$ actually realizes $T^2=-1$, the non-trivial $\Z_2^T$ SPT! Therefore, due to the fractionalization of the spontaneously broken symmetry, the two vacua are in different $\Z_2^T$ SPT phases\footnote{In field theory language, this can be understood as follows. Suppose we start from a theory with $\Z_4^T$ symmetry written as the extension \eqref{eq:Z4T_ext}. Denote by $a \in H^2(M_2,\Z_2)$ the background gauge field for the unitary $\Z_2^m$ subgroup of $\Z_4^T$ and by $w_1 \in H^2(M_2,\Z_2)$ the background for $\Z_2^T$. The relation between the two is $\delta a = w_1 \cup w_1$. Now let us gauge the $\Z_2^m$ subgroup by summing over $a$. The background $\hat{a}$ of the dual $\Z_2^e$ symmetry couples to the original $a$ via a term $\pi i\int_{M_2} \hat{a}\cup a$, but due to $a$ being non-closed, this has a bulk dependency $\nu = \pi i \int_{M_3} \hat{a} \cup w_1 \cup w_1$. Therefore we obtain a theory that has $\Z_2^e \times \Z_2^T$ symmetry with a mixed anomaly $\nu$ \cite{Tachikawa:2017gyf}. The form of this mixed-anomaly implies that whenever $\Z_2^e$ is spontaneously broken, the two vacua must differ by a relative $\Z_2^T$ SPT with action $\pi i \int_{M_2} w_1 \cup w_1$ (see e.g.\ \cite{Debray:2023ior}).}.
    
    \noindent If we choose the other boundary, $\Bphys = \cL_{m,-}$, we obtain the same  $\Z_2^e$ SSB phase stacked with an SPT for the unbroken $\Z_2^T$ subgroup. Therefore, the only difference is that now $v_+$ carries the non-trivial $\Z_2^T$ SPT, while $v_-$ carries the trivial SPT, so that this phase is physically equivalent to the previous one;
    \item[2.] $\Bphys = \cL_{m,\pm}'$: if we choose $\Bphys = \cL_{m,+}'$ we obtain again two vacua $v_\pm = \frac{1\pm \cO_m}{2}$. Since the $m$ anyon is charged under $\Z_2^T$, $v_\pm$ are permuted by both $e$ and $T$, $e(v_\pm) = v_\mp$ and $T(v_\pm) = v_\mp$, but not by the diagonal $eT$ symmetry. The two vacua are still in different $\Z_2^T$ SPT phases, but now for the unbroken $\Z_2^{eT}$ symmetry. Choosing $\Bphys = \cL_{m,-}'$ gives rise again to a physically equivalent phase, where the role of the vacuum carrying the non-trivial SPT is swapped compared to the previous one. 
\end{itemize}

We can obtain the $\Z_4^T$ phases by gauging the internal $\Z_2^e$ subgroup. For phase 1, this leads to the trivially symmetric phase for the $\Z_4^T$ symmetry. In the second case, we obtain a non-trivial SPT phase for the $\Z_4^T$ symmetry. Recall that in phase 2 we have an order parameter $\cO_m$ charged under $e$ and $T$. After gauging $\Z_2^e$, $\cO_m$ is not gauge-invariant, and therefore will be in the twisted sector of the dual symmetry $m$. As $\cO_m$ was originally charged under $T$, we obtain a string order parameter with endpoint charged under $\Z_2^T$. Moreover, notice that now $T^2=m$. Therefore, this type of SPT will be detected by the Klein bottle invariant $\kappa(T,T^2)$, in agreement with the MPS analysis in \cref{sec:klein}. This is also compatible with the known classification $H^2_\epsilon(\Z_4^T,U(1)) = \Z_2$.

\section{Outlook}\label{sec:outlook}
In this work we have considered lattice and field theory approaches to the classification of (1+1)d phases protected by anti-unitary symmetries. The aim was to broaden our understanding of the symmetry-enriched SymTFT approach in the presence of time-reversal symmetry. We analyzed the conditions needed for consistent gapped boundaries of this SymTFT and showed how we can recover the classification of phases that do not break time-reversal symmetry on the boundary. We discussed the subtle lattice string-order parameter calculations in the presence of time-reversal and showed that we can recover the Klein bottle SPT invariant by defining a particular end-point charge. This charge agrees with the charge under the time-reversal action when the end-point is hermitian. In the case of $\Z_2\times \Z_2^T$ we can always choose such a hermitian end-point on the lattice, and we show that this is consistent with the corresponding SymTFT calculation. 

While the framework of space-time symmetry-enriched SymTFT is outlined in Ref.~\cite{Pace:2025hpb}, there is much left to explore in terms of understanding the gapped boundaries of such SymTFTs in different cases. For example, in this work we have considered situations where the $M$-symbol for the twist defects reduces trivially to something depending only on the $G$-enriching symmetry labels. Indeed, we considered cases featuring an abelian anyon $1_g$ in the non-trivial grade,  such that all other anyons in that grade are generated via $a_g = a \times 1_g$ with $a \in \cZ(\cC_1)$. For more general symmetry enrichments, the structure will be more complicated. Nevertheless, since the boundary discussion is formulated entirely in categorical terms (via $M$-symbols and a twisted pentagon relation), it is not tied to group-like symmetries and should apply equally to non-invertible symmetries described by a fusion category. Establishing the framework carefully in the simpler setting of invertible symmetries, as we have done here, is a natural first step towards this, as well as towards the extension to higher-dimensional systems with time-reversal. For recent developments in the direction of non-invertible symmetries see e.g.\ \cite{Wen26}.
Remaining in the $G$-enriched SymTFT framework, it is also not fully clear what would be the implications of using non-symmetric Lagrangian algebras in the sandwich construction\footnote{This happens, for example, in an Ising-type $\Z_2^T$ extension of $\Z_2$, with $T^2 = 1\oplus m$. In the corresponding enriched toric code SymTFT, the $\Z_2^T$ symmetry swaps $e$ and $m$, so clearly there is no gapped boundary that does not explicitly break the enriching symmetry. However, as we have seen, there are also boundaries that break the enriching symmetry only spontaneously. }.

Going beyond time-reversal symmetry, it is natural to consider also parity and CPT symmetry in this context. Indeed, the low-energy (unitary) TQFT has an emergent $\Z_2^{\mathrm{CPT}}$ symmetry \cite{Kapustin:2015uma}. It would be interesting to compare our analysis above to the case of parity symmetry, which acts linearly rather than anti-linearly (and is toggled by the CPT symmetry). Nevertheless we should expect largely similar results (intuitively parity `transposes' and for hermitian operators this will look like time-reversal). With respect to lattice string order parameters, one must define charges carefully also in this case; see the discussion of a string order for $\Z_2^{\mathrm{CPT}}$ in Ref.~\cite{Jones24}. It would be interesting to see how this plays out in the SymTFT. Another way to give time-reversal a unitary action is to take a doubled state and act as a swap\footnote{This is a transposition of the hermitian density matrix, where the doubled state is the Choi state \cite{Ma25}.}. This approach gives rise to the usual `discrete path integral' MPS order parameters that can be used to extract, for example, the crosscap invariant \cite{Pollmann12,Shiozaki16}.  Moreover recent work on applying the SymTFT to classifying mixed state SPTs would be useful here \cite{Ma25,Qi25,Schafer-Nameki:2025fiy,Luo:2025phx}, and would give a different approach to understanding order parameters from a SymTFT perspective.

Beyond this, it would be most interesting to find a broader framework that incorporates gauging of the $\Z_2^T$ and other space-time symmetries. See \cite{Harlow:2023hjb,Susskind:2026moe} for discussion about gauging space-time symmetries in a quantum gravity context and \cite{Apruzzi:2025hvs} for a SymTFT approach in the case of continuous space-time symmetries. 

A less ambitious question is how these ideas apply to higher-dimensional systems, and comparison to lattice approaches and theoretical results \cite{Kapustin:2015uma} (a SymTFT classification for gapped phases in $(2+1)$d has been developed e.g.\ in \cite{Bhardwaj:2024qiv,Bhardwaj:2025piv,Wen:2025thg}). The SymTFT framework has also been applied to the study of gapless theories \cite{Chatterjee:2022tyg,Chatterjee:2022jll,Huang:2023pyk,Bhardwaj:2023bbf,Bhardwaj:2024qrf,Wen:2025thg,Wen:2023otf,Bhardwaj:2025jtf} and it would be interesting to see if this approach can be extended to time-reversal symmetries as well. Note that anti-unitary symmetries have interesting interplay with boundary conformal field theory. For example, they lead to algebraical splitting of edge-modes in time-reversal-enriched Ising conformal field theory \cite{Verresen21}.

\vspace{-.25cm}

\section*{Acknowledgements}
We thank Clement Delcamp for insightful discussions and comments on the draft. We are grateful to Frank Pollmann, Shinsei Ryu, Ryan Thorngren, Ruben Verresen, Dominic Williamson and Matthew Yu for illuminating discussions and correspondence.  We thank Lakshya Bhardwaj, Sakura Schafer-Nameki and Apoorv Tiwari for helpful discussions at a preliminary stage of this project, and Sakura Schafer-Nameki and Rui Wen for coordinating submission of their work \cite{Wen26}. 
\appendix
\vspace{-.25cm}

\section{Lattice string order parameters with an anti-unitary string}\label{app:anti-unitarystring}
Given the analysis of the relation between the Klein bottle invariant and charged end-points of a unitary string in \Cref{sec:klein}, one might wonder if we can reverse the roles of the symmetries and study charged end-points of an anti-unitary string. The anti-unitary string is not an observable in the usual sense. However, this is mathematically well-defined when the state is an MPS; see also the discussion in Ref.~\cite{Chen15}. 
\begin{figure*}
    \centering
    \includegraphics[width=\textwidth]{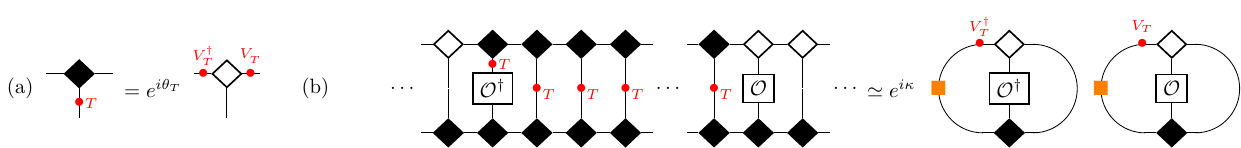}
    \caption{Analogue of \Cref{fig:unitarySO} in the case of an anti-unitary symmetry string.}
    \label{fig:anti-unitarySO}
\end{figure*}\vspace{-.4cm}

\subsection{Anti-unitary string order with end-point charged under a unitary symmetry}
 Let us first consider a string operator for $T$, with end-point charged under some unitary symmetry $h$ (the charge is defined below).
Following the graphical reasoning of \Cref{fig:anti-unitarySO}, we end up with a product of tensor networks of the form
\begin{align}
\rho_R &= \sum \Lambda_\alpha^2 {V}_{T}^{\alpha,\gamma}\overline{\mathcal{A}}_k^{\alpha,\beta} {\mathcal{O}}_{k,j}\mathcal{A}_j^{\gamma,\beta} \qquad \qquad  \rho_L = \sum \Lambda_\alpha^2 (V_T^\dagger)^{\alpha,\gamma}\overline{\mathcal{A}}_k^{\alpha,\beta} \mathcal{O}_{k,j}^\dagger\mathcal{A}_j^{\gamma,\beta}\ .
\end{align}

Since we are studying charge under a unitary $h$, we can directly repeat the analysis of \Cref{fig:unitarySO} and find
\begin{align}
\rho_R&= \sum \Lambda_\alpha^2 (V_h^{} V_T^{} V_h^{\dagger})^{\alpha,\gamma}\overline{\mathcal{A}}_k^{\alpha,\beta} \tilde{\mathcal{O}}_{k,j}\mathcal{A}_j^{\gamma,\beta}\ ,
\end{align}
where $\tilde{\mathcal{O}}_{k,j} = u^{}(h)\mathcal{O} u(h)^\dagger $. However, the combination $V_h^{} V_T^{} V_h^\dagger$ does not have a clear (invariant) representation-theoretic meaning. Indeed, even if $[h,T]=0$, this need not be proportional to $V_T$. Hence, even with a well defined charge $\tilde{\mathcal{O}} = e^{i \chi_{g,h}^\mathcal{O}}{\mathcal{O}}$, we do not learn a gauge invariant phase from the non-vanishing order parameter.

\subsection{Anti-unitary string order with end-point `charged' under anti-unitary symmetry}
Suppose we have an anti-unitary symmetry satisfying $T^2=1$ (note this can include the case of a $\mathbb{Z}_{4k+2}^{\tilde{T}}$ symmetry where $T=\tilde{T}^{2k+1}$).
Using \Cref{fig:anti-unitarySO} as before, we have that the string order is given by a product of $\rho_R$ and $\rho_L$. Then, following the analysis of \cref{sec:klein}
\begin{align}
\rho_R &= \sum \Lambda_\alpha^2 V_T^{\alpha,\gamma}\overline{\mathcal{A}}_k^{\alpha,\beta} \mathcal{O}_{k,j}\mathcal{A}_j^{\gamma,\beta}\nonumber\\
&=  \sum \Lambda_\alpha^2 (V_T V_T^t V_T^\dagger)^{\alpha,\gamma} \overline{\mathcal{A}}_{k}^{\alpha,\beta} \tilde{\mathcal{O}}^t_{k,j} {\mathcal{A}}_j^{\gamma,\beta}\nonumber\\
&=  \theta(T)\sum \Lambda_\alpha^2 V_T^{\alpha,\gamma} \overline{\mathcal{A}}_{k}^{\alpha,\beta} \tilde{\mathcal{O}}^t_{k,j} {\mathcal{A}}_j^{\gamma,\beta}.
\end{align}
Similarly 
\begin{align}
\rho_L &= \sum \Lambda_\alpha^2 (V_T^\dagger)^{\alpha,\gamma}\overline{\mathcal{A}}_k^{\alpha,\beta} \mathcal{O}^\dagger_{k,j}\mathcal{A}_j^{\gamma,\beta}=  \theta(T)\sum \Lambda_\alpha^2 (V_T^\dagger)^{\alpha,\gamma} \overline{\mathcal{A}}_{k}^{\alpha,\beta} (\tilde{\mathcal{O}}^\dagger)^t_{k,j} {\mathcal{A}}_j^{\gamma,\beta}.
\end{align}
Note $\rho_R=\overline{\rho_L}$.
Now, if  $ \tilde{\mathcal{O}}^t=u_T^{} \mathcal{O}^t u_T^\dagger = e^{i \phi_{T,T}^\mathcal{O}} \mathcal{O}$ we get vanishing string order unless $e^{i \phi_{T,T}^\mathcal{O}}=\theta(T)=\pm1$. Fixing $\mathcal{O}$, we have
$ (\tilde{\mathcal{O}}^\dagger)^t = e^{-i \phi_{T,T}^\mathcal{O}} \mathcal{O}^\dagger =e^{i \phi_{T,T}^\mathcal{O}} \mathcal{O}^\dagger $; i.e. the same sign appears.
If $\mathcal{O}$ is hermitian, then this sign is the naive charge under $T$ of the end-point.

In fact, since $T^2=1$, we can always find a hermitian end-point operator. To see this, suppose that we find some end-point operator $\mathcal{O}$ with long-range order. Then one can show that at least one of $(\mathcal{O}+\mathcal{O}^\dagger)$ and $i(\mathcal{O}-\mathcal{O}^\dagger)$ have long-range order \cite{Verresen21}.  

We can thus identify the crosscap invariant from the charged end-point of an anti-unitary string. This should be contrasted with approaches based on taking two copies of the state to construct an analogue of the path integral on $\mathbb{R}P^2$ \cite{Pollmann12,Shiozaki16}.
\vspace{-.3cm}
\bibliography{article.bbl}
\end{document}